\documentclass[aps,showpacs]{revtex4}
\usepackage[utf8]{inputenc}
\usepackage{color}
\usepackage{amsmath}
\usepackage{graphicx}

\makeatletter
\@ifundefined{textcolor}{}
{%
 \definecolor{BLACK}{gray}{0}
 \definecolor{WHITE}{gray}{1}
 \definecolor{RED}{rgb}{1,0,0}
 \definecolor{GREEN}{rgb}{0,1,0}
 \definecolor{BLUE}{rgb}{0,0,1}
 \definecolor{CYAN}{cmyk}{1,0,0,0}
 \definecolor{MAGENTA}{cmyk}{0,1,0,0}
 \definecolor{YELLOW}{cmyk}{0,0,1,0}
}

\usepackage{hyperref}
\usepackage{dcolumn}
\usepackage{bm}
\usepackage{wrapfig}
\usepackage{subfigure}
\usepackage{ucs}
\usepackage{lineno}
\usepackage{epsfig}
\usepackage{psfrag}\usepackage{epstopdf}
\DeclareGraphicsExtensions{.pdf,.eps,.png,.jpg,.mps}

\makeatother
\begin{document}

% \preprint{APS/123-QED}

\title{Spin-1/2 Ising-Heisenberg Cairo pentagonal model in the presence of an external magnetic field: Effect of  Land{\'e} g-factors}
 
\author{Hamid Arian Zad }
\email{arianzad.hamid@yerphi.am}
 \affiliation{ A. I. Alikhanyan National Science Laboratory, Alikhanian Br. 2, 0036 Yerevan, Armenia}%Lines break automatically or can be forced with \\
 \affiliation{ ICTP, Strada Costiera 11, I-34151 Trieste, Italy}
% \email{arianzad.hamid@mshdiau.ac.ir}

\author{Andrea Trombettoni}
\email{andreatr@sissa.it}
 \affiliation{ Department of Physics, University of Trieste, Strada Costiera 11, I-34151 Trieste, Italy}
\affiliation{ CNR-IOM DEMOCRITOS Simulation Center and SISSA, Via Bonomea 265, I-34136 Trieste, Italy}%

\author{Nerses Ananikian }
\email{ananik@yerphi.am}
 \affiliation{A. I. Alikhanyan National Science Laboratory, Alikhanian Br. 2, 0036 Yerevan, Armenia}
 \affiliation{ CANDLE Synchrotron Research Institute, Acharyan 31, 0040 Yerevan, Armenia}
\affiliation{ BLTP, Joint Institute for Nuclear Research, Dubna, Moscow Region 141980,  Russia}%




\begin{abstract}
 In the present paper, a study of the magnetic properties
  of a spin-1/2 Ising-Heisenberg Cairo pentagonal structure is %reported
  presented.
  The model has been investigated 
  in Ref. \cite{Rodrigues} in the absence of external magnetic field.
  Here, 
  we consider the effects
  %in the presence
  of an external tunable magnetic field.
  By using the transfer matrix approach, we
  investigate the magnetic ground-state phase transition, the low-temperature
  magnetization process, and how the magnetic field influences
  the various thermodynamic parameters %of the model
  such as entropy, internal energy and specific heat.
  It is shown that the model exhibits intermediate magnetization plateaux
  %at zero, one-fourth and one-half of the saturation magnetization
  accompanied by %growing
  a double-peak in the specific heat curve versus temperature.
  The position of each magnetization jump is in accordance with the 
  merging and/or separation %each
  of the two peaks in the specific heat curve.
  Considering different g-factors for the nodal Ising spins
  and spin dimers also results in arising different intermediate
  plateaux and to remarkable alterations of the thermodynamic properties of the model.
\end{abstract}

\maketitle

%\tableofcontents

\section{Introduction}
The study of 
%about interacting quantum many-body systems like
 quantum spin systems with competing interactions is an active field of research in solid state physics.
%a number of subjects such as
%In these particular areas
 Efforts have focused on ferrimagnetic chains, since
       they feature both ferromagnetic and antiferromagnetic phases,
       which have been studied both at zero and finite temperature      
\cite{Kitaev,Gu,Dillenschneider,Ivanov,Werlang,Sachdev,Rojas2,RojasM,Strecka}.
Spin ladders %can be count as attractive solveable models among these systems.
have been also extensively studied
using various methods \cite{Strecka,Koga,Okamoto,Muller,Vuletic,Notbohm,Buttner,Blundell,Bacq,Arian1,Feng2007,Giuseppe2020}.
Analogously, 2-D lattices like spin-1/2 Heisenberg antiferromagnets
on the kagome lattice have been also widely examined within the Lanczos
diagonalization both for the
  ground state properties up to $48$ spins \cite{Moessner2019}
  and at finite temperature up to $42$ spins \cite{Schnack2018},
   and using DMRG \cite{White2011,Kolley2015}.

Along the years, the synthesis of
compounds such as $\mathrm{A}_3\mathrm{Cu}_3(\mathrm{PO}_4)_4$
with $\mathrm{A}=\mathrm{Ca}, \mathrm{Sr}$ \cite{Drillon1},
$\mathrm{Cu}_3\mathrm{Cl}_6(\mathrm{H}_2\mathrm{O})_2 \\
\cdot 2\mathrm{H}_8\mathrm{C}_4\mathrm{SO}_2$ \cite{Okamoto1999,Okamoto2003},
the diamond chain like  polymers  $\mathrm{Cu}_3(\mathrm{TeO}_3)_2\mathrm{Br}_2$ \cite{Uematsu}
and the natural mineral azurite  $[\mathrm{Cu}_3(\mathrm{CO}_3)_2(\mathrm{OH})_2]$ \cite{Kikuchi1,Kikuchi2},
 became possible. These materials can be
studied in terms of Heisenberg spin models.
%Recently, A. Baniodeh {\it et al.} verified experimentally the ground-state as well as low-temperature thermodynamic properties of material \\
%$\big[\mathrm{Fe}_{10}\mathrm{Gd}_{10}(\mathrm{Me-tea})_{10}(\mathrm{Me-teaH})_{10}(\mathrm{NO}_3)_{10}\big]\cdot 20\mathrm{MeCN}$
Recently, a mixed $3d/4f$ cyclic coordination cluster with
a ground-state spin of $S = 60$ was studied experimentally in
\cite{Baniodeh}, where the authors
synthesised the $\mathrm{Gd}$-containing isotropic member of a
new series of cyclic coordination clusters, forming a nano-torus with
alternating gadolinium and iron ions with a nearest neighbour
$\mathrm{Fe}–\mathrm{Gd}$
coupling and a frustrating next-nearest neighbour $\mathrm{Fe}–\mathrm{Fe}$
coupling. This arrangement corresponds to a cyclic delta or saw-tooth chain
\cite{Baniodeh}.
Motivated by the compound $\mathrm{Bi}_2\mathrm{Fe}_4\mathrm{O}_9$,
in Ref. \cite{Rojas2012} %, M. Rojas {\it et al.} %offered gave
it was given a general solution for the frustrated Ising model
on the so-called 2-D Cairo pentagonal lattice, i.e. a planar lattice
where the tiling is achieved with non-regular pentagons.
%; the lattice may be viewed as an assembly of checkerboard ordering
%with the elementary cell (see Fig. 5) rotated by π/2 in the
%neighboring square plaquettes, as shown in Fig. 1 (f 
In Ref. \cite{Rousochatzakis2012} %authors
the antiferromagnetic Heisenberg model on the Cairo pentagonal
lattice was studied.
Afterwards, F. C. Rodrigues {\it et al.} sketched a stripe of the Cairo pentagonal Ising-Heisenberg model,
and studied the ground-state phase transition and some
thermodynamic parameters for such model in \cite{Rodrigues}.
The advantage of the Cairo pentagonal Ising-Heisenberg stripe 
geometry is to make possible
analytical calculations, and also the possibility
of considering the effect of the interactions
of classical spins interacting with quantum ones. Another advantage, exploited
in the following, is that it allows for an evaluation of the effect of
external magnetic fields.
 
Phase transitions in spin models with competing interactions
has been one of the most interesting
topics of condensed matter physics
and statistical mechanics during the last decades 
\cite{Arian1,Feng2007,Saadatmand,Valverde,Ananikian2012,Abgaryan1,Strecka1,Arian2,Campa20}.
Further investigations of the spin models in the presence of
an external magnetic field have supplied
exact outcomes for the ground-state phase transition,
which can be induced through the exchange couplings
\cite{Sahoo1,Sahoo2,Giri,Hovhannisyan}. %It is quite noteworthy that
The ground-state of the spin ladders with
 higher spins have been examined so far \cite{Ivanov,Giri}.
%as well 

The investigation of magnetization curves and plateaux has attracted
 considerable interest.
Exactly solvable quantum spin models for which magnetization varies
smoothly as a function of the magnetic field until
reaches its saturation magnetization include
spin-1/2 quantum chains 
\cite{Hida,Verkholyak}, ladders \cite{Arian1,Cabra}
and  spin-1/2 Ising-Heisenberg diamond chains
\cite{Gu,Rojas2,RojasM,Ananikian2012,Abgaryan1,Strecka1}. 
%For the small quantum spin clusters, V. Ohanyan {\it et al.} investigated general non-commutativity features of the magnetization operator and Hamiltonian \cite{Ohanyan2015}.
The specific heat of magnetic materials
%, as an applicable topic in statistical physics,
has attracted as well much attention, %over the past two decades,
since it %usually
may exhibit an anomalous thermal behavior when %by altering other
the parameters of the Hamiltonian such as coupling constants, spin exchange anisotropy and magnetic field are changed.
Such a function can be estimated by the Schottky theory \cite{Strecka,Karlova2016}.
The associated round maximum (Schottky maximum) appeared in the specific heat curve, has been detected in various magnetic materials in experiments \cite{Bernu,Misguich,Helton}.

%In solid state physics and material science, most of the theoretical
%treatments are based on numerical techniques. Hence,
When practicable, the possibility of using an analytical approach to
describe the ground-state diagram and magnetic and thermodynamic properties
of quantum spin systems such as magnetization, entropy, internal energy and
specific heat, is %definitely required
certainly unvaluable. A very useful %promising
method is the transfer-matrix formalism which has widely been applied
to a number of strongly correlated systems at zero-temperature and
%, as well as,
low temperature for studying the ground- and low-lying state properties
of spin models. In the following of the analytical discussion in
Ref. \cite{Rodrigues}, we here investigate the magnetic and
thermodynamic properties of the spin-1/2 Ising-Heisenberg Cairo
pentagonal model in the presence of an external magnetic field
using the same transfer matrix technique. We also consider %a unique situation
the possibility to have in this model
%for the model such that
different Land{\' e} g-factors \cite{Vadim2015,Vadim2020}, showing
that the diffence among them plays an important role in the
magnetic and thermodynamic behaviors of the model.
%In spite of its matter, the later situation has not  been considered for the spin-1/2 Ising-Heisenberg Cairo pentagonal model yet.

\begin{figure*}[t!]
  \centering
  \resizebox{1\textwidth}{!}{
 \includegraphics[trim=10 60 10 60, clip]{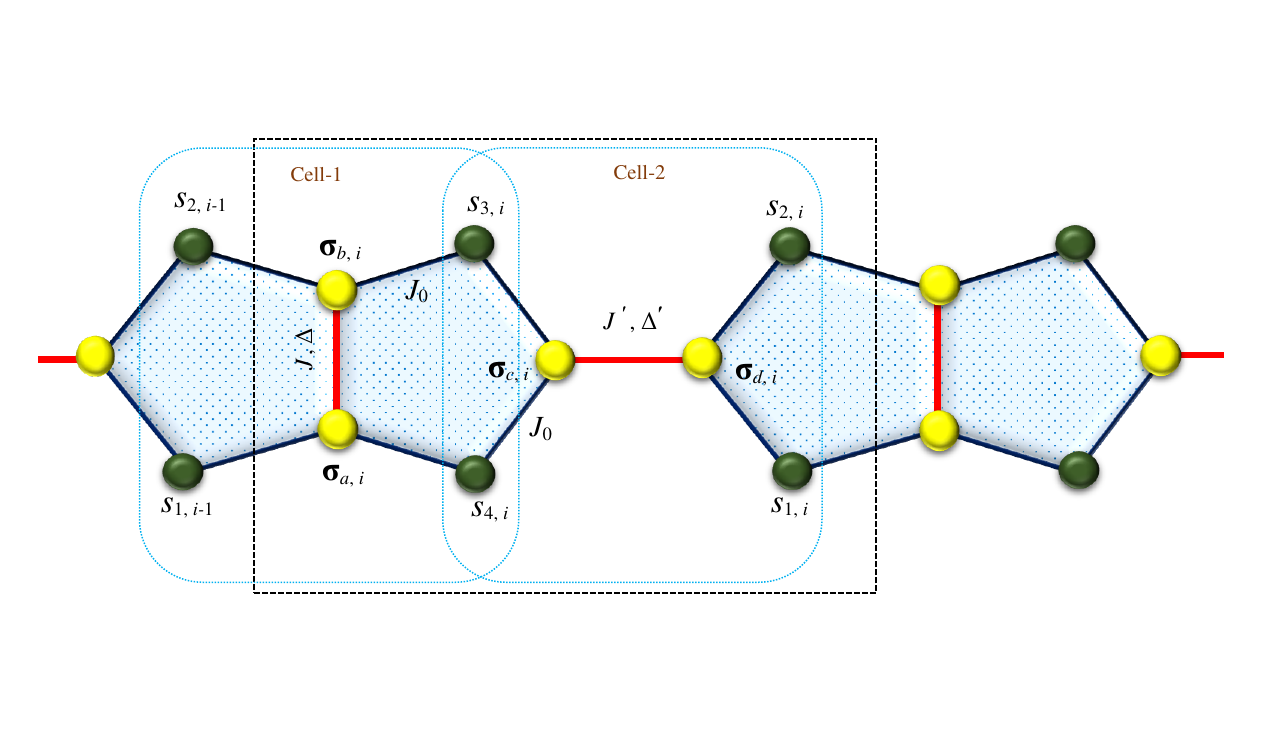} %[trim=left bottom right top, clip] 
}
  \caption{Schematic representation of the spin-1/2 Ising-Heisenberg Cairo pentagonal  model. Yellow %balls
    circles represent Heisenberg dimer spins and
    green %balls
    circles between them, localized on the %wings
     external vertices of each pentagon, represent
    the Ising nodal spins. The dashed rectangle represents an unit block.}
\label{fig:SpinLadder}
\end{figure*}

The paper is organized as follows. In Sec. \ref{Model} we describe the
%exactly solvable
model and present its thermodynamic solution with in
the transfer-matrix formalism. In Sec. \ref{TP},
we numerically discuss the magnetization process and
the thermodynamic behavior of the model in the presence
of an external homogeneous magnetic field. Finally, the most significant
results will be
summarized together in Sec. \ref{conclusions}.

\section{Model and exact solution within the transfear matrix formalism}\label{Model}
The Hamiltonian of the spin-1/2 Ising-Heisenberg Cairo pentagonal model shown in Fig. \ref{fig:SpinLadder} can be written as
\begin{equation}\label{Hamiltonian}
\begin{array}{lcl}
  H = \sum\limits_{i=1}^N \big[\mathcal{H}_{i-1,i}^{ab}+\mathcal{H}_{i, i}^{cd}\big] +
  %\mathcal{H}_z,
  {H_z},
\end{array}
\end{equation}
where the Hamiltonians of two sub-units block cell-1 ($\mathcal{H}_{i-1,i}^{ab}$) and cell-2 ($\mathcal{H}_{i, i}^{cd}$)
%lied
in each block together with the Zeeman term
%effect
are given by
\begin{equation}\label{block Hamiltonian}
\begin{array}{lcl}
\mathcal{H}_{i-1,i}^{ab}= -J ({\boldsymbol \sigma}_{a,i}\cdot{\boldsymbol \sigma}_{b,i})_{\Delta}-J_{0} \big(s_{1,i-1}  +  s_{4,i}\big){\sigma}_{a,i}^z  - J_{0} \big(s_{2,i-1}+ s_{3,i}) \sigma_{b,i}^z, \\
\mathcal{H}_{i, i}^{cd}= -J^{\prime} ({\boldsymbol \sigma}_{c,i}\cdot{\boldsymbol \sigma}_{d,i})_{\Delta^{\prime}}- J_{0} \big(s_{3,i}  +  s_{4,i}\big){\sigma}_{c,i}^z 
- J_{0} \big(s_{1,i}+ s_{2,i}) \sigma_{d,i}^z, \\
H_z = -\mu_B B_z\sum\limits_{i=1}^{N} {\big[} g_1{\big(}
  \sigma_{a,i}^z+\sigma_{b,i}^z+\sigma_{c,i}^z+\sigma_{d,i}^z{\big)}
+\dfrac{1}{2}g_2{\big(} s_{1,i}+s_{2,i}+s_{3,i}+s_{4,i}
{\big) \big]},
\end{array}
\end{equation}
where $N$ is the number of unit blocks. Two different Land{\'e} g-factors
$g_1$ and $g_2$ are considered. The XXZ interaction between pair
spins of $ab$-dimer can be given by  
\begin{equation}
\begin{array}{lcl}
J ({\boldsymbol \sigma}_{a,i}\cdot{\boldsymbol \sigma}_{b,i})_{\Delta}\equiv J\big(\sigma_{a,i}^x\sigma_{b,i}^x+\sigma_{a,i}^y\sigma_{b,i}^y\big)+\Delta \sigma_{a,i}^z\sigma_{b,i}^z.
\end{array}
\end{equation}
Analogously, for the $cd$-dimer we have following definition
\begin{equation}
\begin{array}{lcl}
J^{\prime} ({\boldsymbol \sigma}_{c,i}\cdot{\boldsymbol \sigma}_{d,i})_{\Delta^{\prime}}\equiv J^{\prime}\big(\sigma_{c,i}^x\sigma_{d,i}^x+\sigma_{c,i}^y\sigma_{d,i}^y\big)+\Delta^{\prime} \sigma_{c,i}^z\sigma_{d,i}^z.
\end{array}
\end{equation}
The nodal spins $s^{\alpha,i}$ ($\alpha=\{1, 2, 3, 4\}$) localized on the wings of  each pentagon are Pauli operators, taking values $(-1,\;1)$. The pure Ising-type exchange coupling $J_0$ represents interaction between nodal spins $s^{\alpha,i}$ and Heisenberg dimer spins.
 $\boldsymbol{\sigma}=\lbrace {\sigma}^x, {\sigma}^y, {\sigma}^z \rbrace$ are Pauli operators of Heisenberg dimers (with $\hbar=1$) and are given by
 \begin{equation*}\label{TM}
\sigma^x= \left(
\begin{array}{cc}
 0 &1 \\
 1 & 0 
\end{array} \right), \quad 
\sigma^y= \left(
\begin{array}{cc}
 0 & -i \\
 i & 0 
\end{array} \right), \quad
\sigma^z= \left(
\begin{array}{cc}
 1 & 0 \\
 0 &  -1 
\end{array} \right).
\end{equation*}
 The final part of the Hamiltonian (\ref{Hamiltonian}) accounts for the
 Zeeman$^,$s energy of magnetic moments in the external magnetic field
 $B={B}_z$. We will write
 {$H_z=\sum_{i=1}^N \mathcal{H}^{i}_z$,
   with $\mathcal{H}^{i}_z = (\mathcal{H}_{i-1,i}^{ab})^z+(\mathcal{H}_{i,i}^{cd})^z$
   representing the Zeeman$^,$s energy of each two cells in the block Hamiltonian, and 
   $H=\sum_{i=1}^N \mathcal{H}_{i}$, with
   $\mathcal{H}_{i}=\mathcal{H}_{i-1,i}^{ab}+ \mathcal{H}_{i,i}^{cd}+\mathcal{H}^{i}_z$.}
From now on, we consider $J_0$ as energy unit
  for all other parameters with $B/J_0$,  $J/J_0$, $\Delta/J_0$ and $T/J_0$ being dimensionless parameters, restoring $J_0$ in equations and plots when needed for clarity. Moreover, we will set $\mu_B=1$.
  % while  $g_1$ and $g_2$ are assumed to be constant values.
 
The commutation relation between each of two different block Hamiltonians
 $[\mathcal{H}_i, \mathcal{H}_j] = 0$ enables us to extract
 the partition function of the model under consideration from the following formula
\begin{equation}\label{PF}
  {Z}=Tr\Big[\displaystyle\prod_{i=1}^{%N^{\prime}
      {N}}\exp\big(-\beta
    %\big[\mathcal{H}_{i-1,i}^{ab}+ \mathcal{H}_{i,i}^{cd}+\mathcal{H}^{i}_z\big]
   {\mathcal{H}_{i}} \big)\Big],
\end{equation}
where %, $\mathcal{H}^{i}_z=(\mathcal{H}_{i-1,i}^{ab})^z+(\mathcal{H}_{i,i}^{cd})^z$
%represents the Zeeman$^,$s energy of each two cells in the block Hamiltonian.
$\beta=1/k_{B}T$ with $k_{B}$ the Boltzmann constant and $T$ the temperature
(for simplicity we set $k_{B}=1$). Hence, we can write the $4\times 4$ transfer matrix
$\mathcal{W}$ as it follows:
\begin{equation}\label{TrM}
\begin{array}{lcl}
\mathcal{W}=\mathcal{T}_{ab}\mathcal{T}_{cd}=\sum\limits_{k=1}^{4}\exp\big[-\beta\mathcal{E}_k(s_{1,i-1}\;s_{2,i-1}\mid s_{4,i}\;s_{3,i})\big]
 \sum\limits_{k=1}^{4}\exp\big[-\beta\bar{\mathcal{E}}_k(s_{1,i}\;s_{4,i}\mid s_{2,i}\;s_{3,i})\big].
\end{array}
\end{equation}
$\mathcal{T}_{ab}$ represents the transfer matrix of the cell-1 and $\mathcal{T}_{cd}$
represents the transfer matrix of the cell-2. 
Symbol $\mid$ distinguishes the interplay between rows (pair nodal spins $\{s_{1,i-1},\;s_{2,i-1}\}$) and the columns (nodal spins 
$\{s_{4,i},\;s_{3,i}\}$) of the transfer matrix $\mathcal{T}_{ab}$, and between rows (pair nodal spins $\{s_{1,i},\;s_{4,i}\}$) and the columns (nodal spins $\{s_{2,i},\;s_{3,i}\}$) of the transfer matrix $\mathcal{T}_{cd}$.
%\textcolor{red}{In %writing
 % Eq. (\ref{TrM}) we stick to the convention of not explicitly writing the contribution
 % of the classical part of the Hamiltonian to the transfer matrix $\mathcal{W}$.}
Moreover, $\mathcal{E}_k$ denote the eigenvalues of the Hamiltonian
$\mathcal{H}_{i-1,i}^{ab}+(\mathcal{H}_{i-1,i}^{ab})^z$ and depend on the
spins $s_{\alpha,i}$ (with $\alpha=1,2,3,4$). One obtains
%\textcolor{red}{
\begin{equation}\label{TrM_1}
\begin{array}{lcl}
%\mathcal{E}_{1,2}=-\frac{\Delta}{4}\pm\frac 14\left(2g_1B+J(s_{1,i-1}+s_{2,i-1}+s_{3,i}+s_{4,i})\right),\\
%\mathcal{E}_{3,4}=\frac{\Delta}{4}\pm\frac 12 \sqrt{J_0^2\left(s_{1,i-1}+s_{4,i}-s_{2,i-1}-s_{3,i}\right)^2+J^2}.
%\mathcal{E}_{1,2}=-\frac{\Delta}{4}\pm\frac 14\left(2g_1+J(s_{1,i-1}+s_{2,i-1}+s_{3,i}+s_{4,i})\right),\\
%\mathcal{E}_{3,4}=\frac{\Delta}{4}\pm\frac 12 \sqrt{J_0^2\left(s_{1,i-1}+s_{4,i}-s_{2,i-1}-s_{3,i}\right)^2+J^2}.
\mathcal{E}_{1,2}=-\frac{\Delta}{4}-\frac{g_2B}{4}(s_{1,i-1}+s_{2,i-1}+s_{3,i}+s_{4,i})\pm\frac 14\left(2g_1B+J(s_{1,i-1}+s_{2,i-1}+s_{3,i}+s_{4,i})\right),\\
\mathcal{E}_{3,4}=\frac{\Delta}{4}-\frac{g_2B}{4}(s_{1,i-1}+s_{2,i-1}+s_{3,i}+s_{4,i}) \pm\frac 12 \sqrt{J_0^2\left(s_{1,i-1}+s_{4,i}-s_{2,i-1}-s_{3,i}\right)^2+J^2}.
\end{array}
\end{equation}
%}
Analogously, the corresponding eigenenergies for $cd$-dimer are given by
%\textcolor{red}{
\begin{equation}\label{TrM_2}
\begin{array}{lcl}
%\bar{\mathcal{E}}_{1,2}=-\frac{\Delta^{\prime}}{4}\pm\frac 14\left(2g_1B+J^{\prime}(s_{1,i}+s_{2,i}+s_{3,i}+s_{4,i})\right),\\
%\bar{\mathcal{E}}_{3,4}=\frac{\Delta^{\prime}}{4}\pm\frac 12 \sqrt{J_0^2\left(s_{1,i}+s_{2,i}-s_{4,i}-s_{3,i}\right)^2+J^{\prime 2}}.
%\bar{\mathcal{E}}_{1,2}=-\frac{\Delta^{\prime}}{4}\pm\frac 14\left(2g_1+J^{\prime}(s_{1,i}+s_{2,i}+s_{3,i}+s_{4,i})\right),\\
%\bar{\mathcal{E}}_{3,4}=\frac{\Delta^{\prime}}{4}\pm\frac 12 \sqrt{J_0^2\left(s_{1,i}+s_{2,i}-s_{4,i}-s_{3,i}\right)^2+J^{\prime 2}}.
\bar{\mathcal{E}}_{1,2}=-\frac{\Delta^{\prime}}{4}-\frac{g_2B}{4}(s_{1,i}+s_{2,i}+s_{3,i}+s_{4,i})\pm\frac 14\left(2g_1B+J^{\prime}(s_{1,i}+s_{2,i}+s_{3,i}+s_{4,i})\right),\\
\bar{\mathcal{E}}_{3,4}=\frac{\Delta^{\prime}}{4}-\frac{g_2B}{4}(s_{1,i}+s_{2,i}+s_{3,i}+s_{4,i})\pm\frac 12 \sqrt{J_0^2\left(s_{1,i}+s_{2,i}-s_{4,i}-s_{3,i}\right)^2+J^{\prime 2}}.
\end{array}
\end{equation}
%}

Hereafter, for simplicity, we consider the %general
case $J^{\prime}=J$ and $\Delta^{\prime}=\Delta$. Since we look for eigenvalues
of the transfer matrix $\mathcal{W}$ in the thermodynamic limit $N\rightarrow\infty$,
as usual the largest eigenvalue $\Lambda_{max}$ is the only one determining
%has the most effect on
the thermodynamic properties of the system \cite{Yeomans}. %, whereas the other three smaller eigenvalues are almost effectless and their contribution can be completely neglected.
Hence, the free energy per block can be obtained from the largest eigenvalue
of the transfer matrix (\ref{TrM}) as
 \begin{equation}\label{FreeE}
 \begin{array}{lcl}
 f=-\frac{1}{\beta}\lim\limits_{N\rightarrow \infty}\ln\frac{1}{N}{Z}=-\frac{1}{\beta}\ln\Lambda_{max}.
 \end{array}
 \end{equation}
The magnetization, entropy, and specific heat per block can be defined as
\begin{equation}\label{TParameters}
\begin{array}{lcl}
{M}=-\Big(\frac{\partial f}{\partial B}\Big)_{T}\; , \; {S}=-\Big(\frac{\partial f}{\partial T}\Big)_{B}\;, \quad {C}=-T\Big(\frac{\partial^2 f}{\partial T^2}\Big)_{B}.
\end{array}
\end{equation}

\section{Results and discussion}\label{TP}

This section introduces the results obtained from the study of the possible ground-state
phase transitions, magnetization process, entropy, internal energy and
specific heat behavior of the spin-1/2 Ising-Heisenberg Cairo pentagonal model
in different cases.

\subsection{Magnetization and ground-state phase diagram}

\begin{figure*}
\begin{center}
\resizebox{0.45\textwidth}{!}{%
\includegraphics[trim=10 0 40 30, clip]{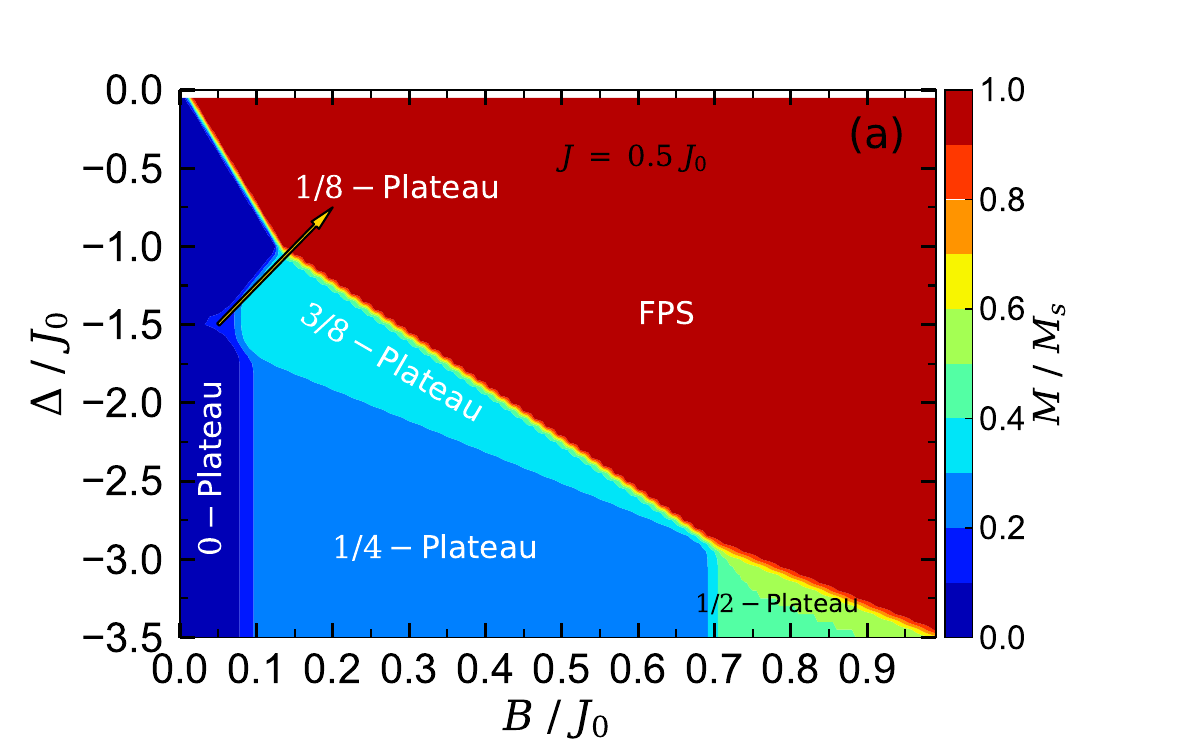}
}
\resizebox{0.45\textwidth}{!}{%
\includegraphics[trim=10 0 40 30, clip]{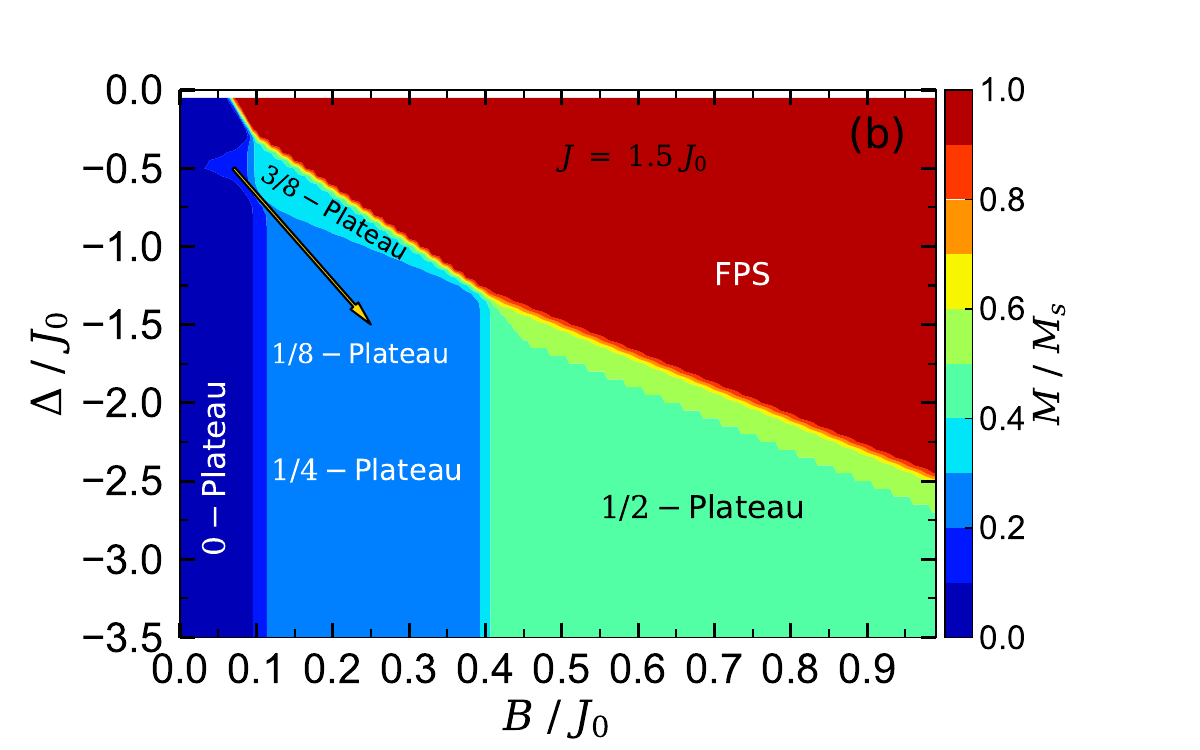}  
}
\resizebox{0.45\textwidth}{!}{%
\includegraphics[trim=10 0 40 30, clip]{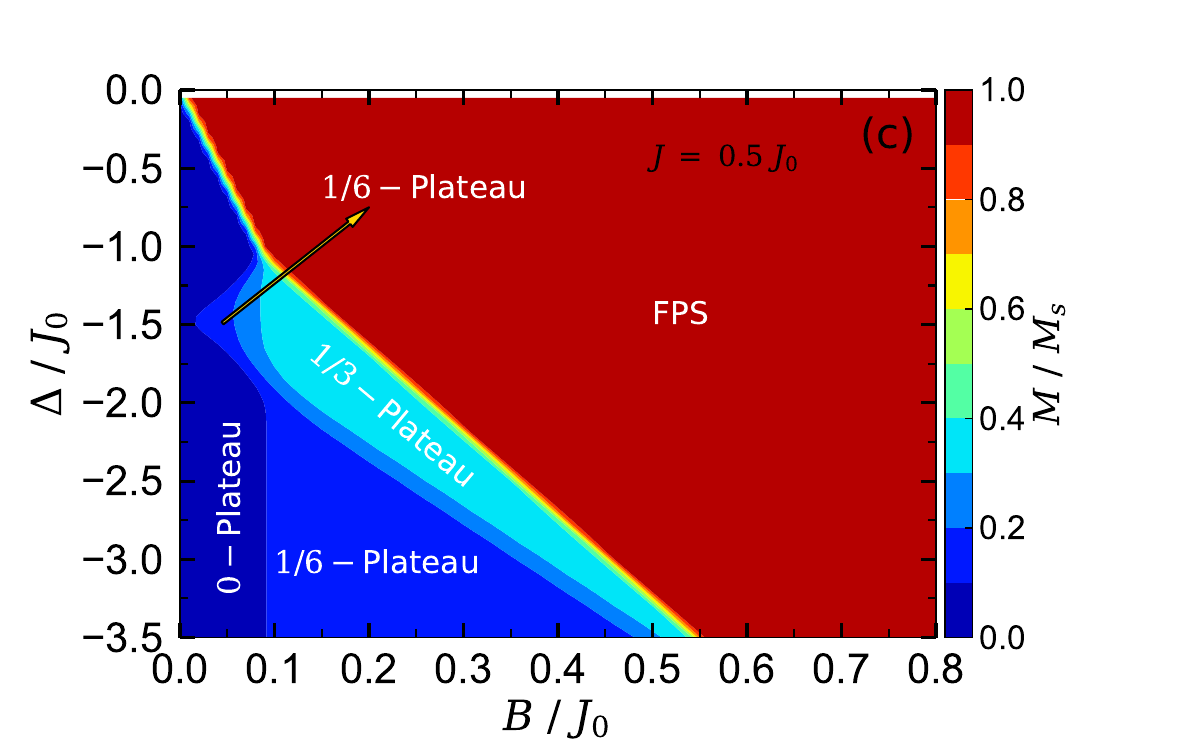}
}
\resizebox{0.45\textwidth}{!}{%
\includegraphics[trim=10 0 40 30, clip]{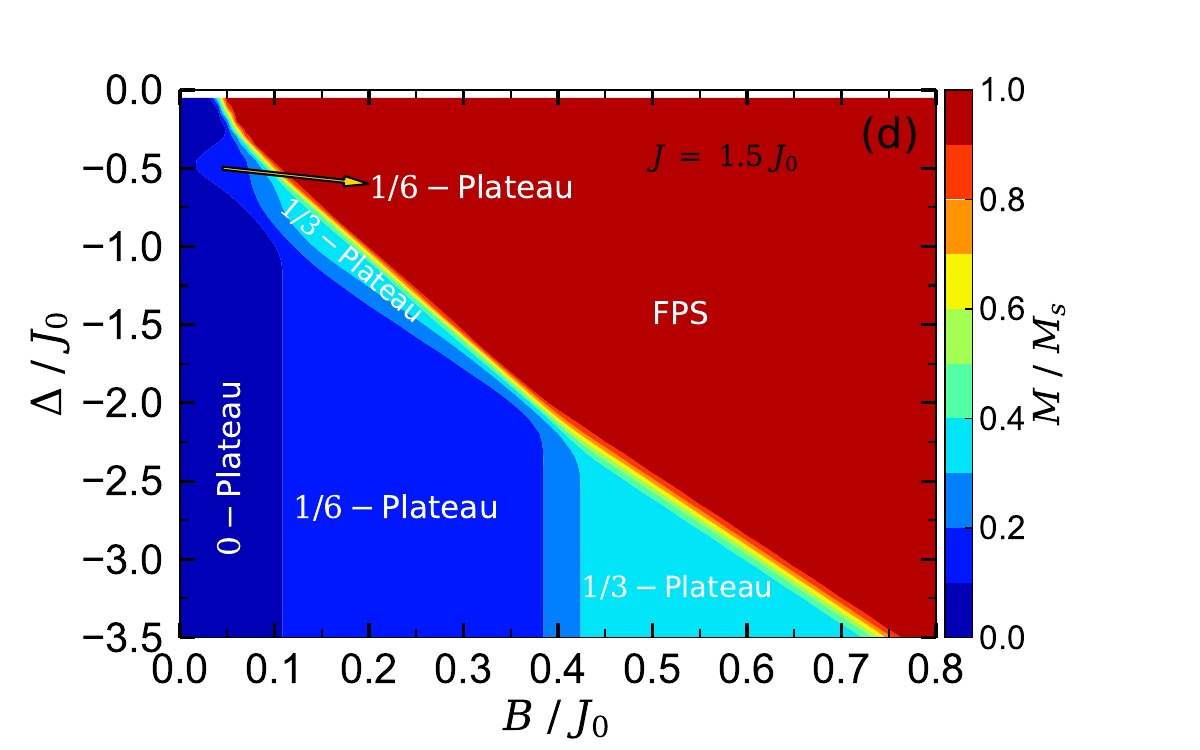}  
}
\caption{The ground-state magnetic phase diagram of the model in the ($B/J_0-\Delta/J_0$)
  plane by considering fixed values: (a) $J=0.5J_{0}$ and $g_1=g_2=1$;
  (b) $J=1.5J_{0}$ and $g_1=g_2=1$; (c) $J=0.5J_{0}$, $g_1=1$ and $g_2=2$;
  (d) $J=1.5J_{0}$, $g_1=1$ and $g_2=2$.}
\label{fig:QPT}
\end{center}
\end{figure*}

{We start by considering the case $g_1=g_2$ for the Land{\' e} g-factors.}
Various possible magnetic ground-states of a unit-block of the spin-1/2 Ising-Heisenberg
Cairo pentagonal model can be identified in terms of six different phases
with the corresponding magnetization plateax at zero, one-eighth, one-fourth,
three-eighth and one-half of saturation magnetization and fully polarized state (FPS).  
The magnetic ground-state phase diagrams in the ($B/J_0-\Delta/J_0$) plane are displayed
in Figs. \ref{fig:QPT}(a) and (b) for fixed values of, respectively, $J =0.5J_0$ and
$J =1.5J_0$ by supposing the same g-factors for the nodal Ising spins and
Heisenberg dimers, i.e., $g_1=1$ and $g_2=1$.
In Fig. \ref{fig:QPT}(a) we see that at anisotropy range $\Delta> -J_0$
there is a single magnetization jump from the zero plateau to the FPS saturation magnetization. When the anisotropy decreases, the magnetization curve reveals intermediate magnetization
plateau at one-eighth of the saturation magnetization at
{%$0.05J_0\leq B< 0.1 J_0$
  $0.05J_0\lesssim B \lesssim 0.1 J_0$}.
Meanwhile, intermediate plateaux at three-eighth and one-forth of the saturation value
are present %illustrated at
in the magnetic field range %$0.1J_0\leq B< 0.7 J_0$
{$0.1J_0\lesssim B \lesssim 0.7 J_0$}
when the anisotropy changes in the interval
%$-3J_0\leq\Delta\leq -J_0$.
{$-3J_0\lesssim\Delta\lesssim -J_0$}.

With a further decrease of the anisotropy
%($\Delta< -3J_0$)
{$\Delta \lesssim -3J_0$}
another intermediate plateau
at one-half of saturation value is %illustrated
observed for the magnetic field range $B\gtrsim 0.7 J_0$.
It is quite visible from Fig. \ref{fig:QPT}(b) that the increment of $J/J_0$
leads to broaden the area of one-half plateau.

{Let now study the case $g_1 \neq g_2$, considering the ratio $g_2/g_1$
  being an integer. 
  %By considering different g-factors for the nodal Ising spins and Heisenberg dimers, i.e.,
  For $g_1=1$ and $g_2=2$}, as in Figs. \ref{fig:QPT}(c) and \ref{fig:QPT}(d), the model has
%goes to
ground-states with completely different magnetization intermediate plateaux at zero,
one-sixth and one-third normalized with respect to its saturation value.

\begin{figure*}[t!]
\begin{center}
\resizebox{0.45\textwidth}{!}{%
\includegraphics{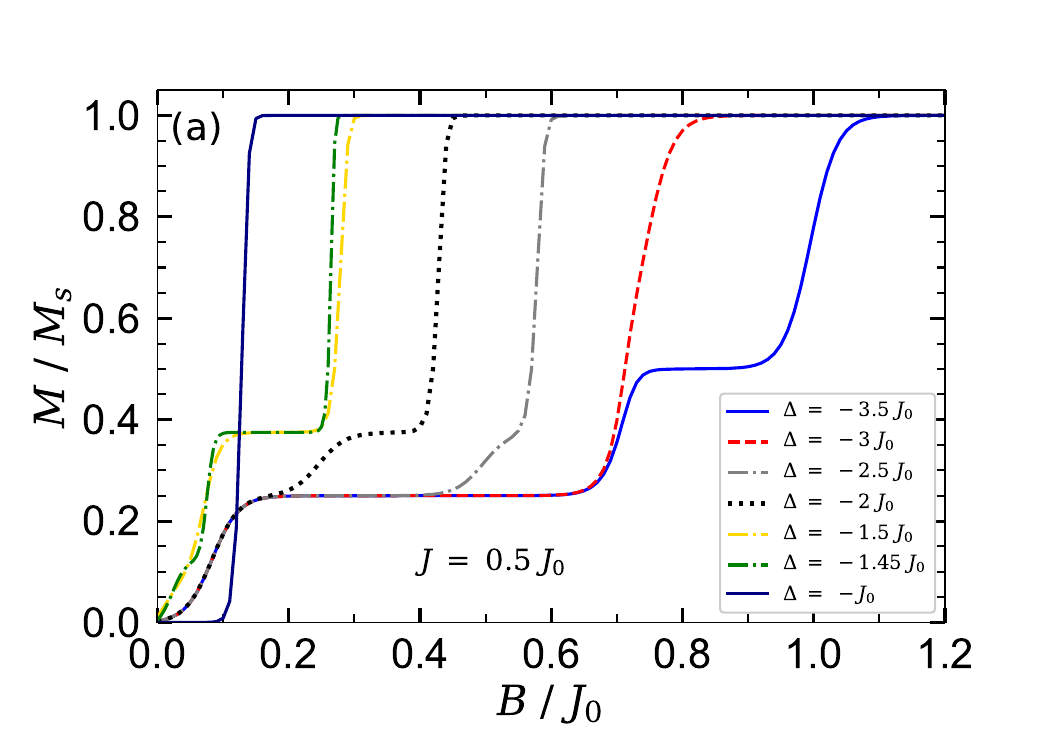}
}
\resizebox{0.45\textwidth}{!}{%
\includegraphics{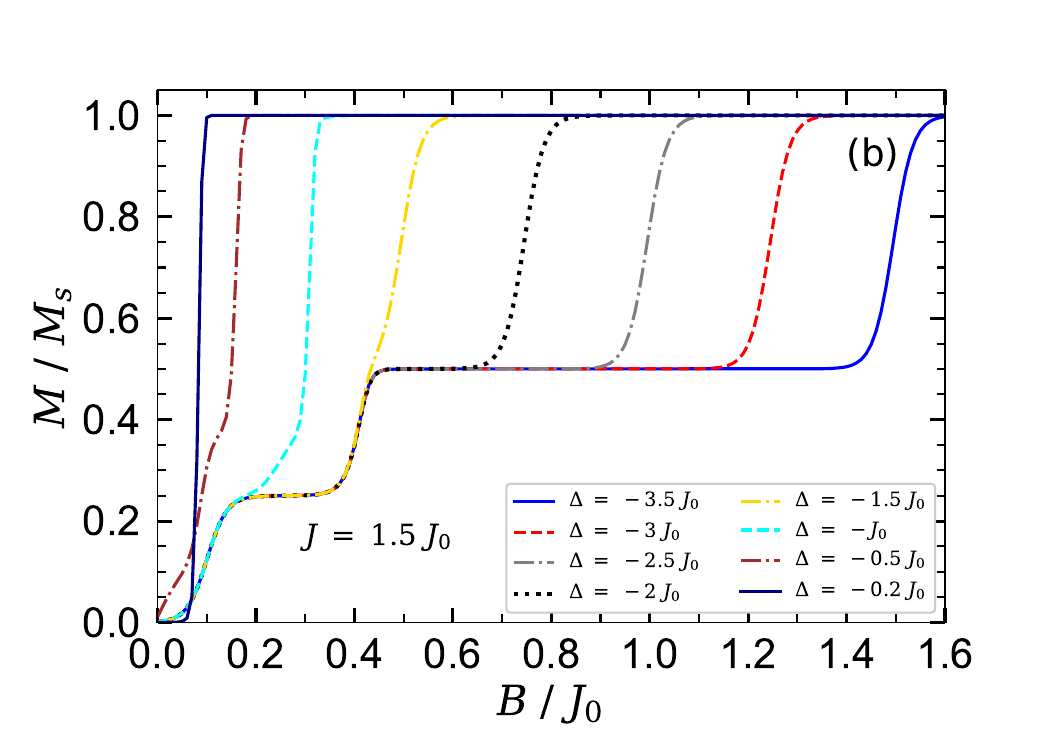}
}
\resizebox{0.45\textwidth}{!}{%
\includegraphics{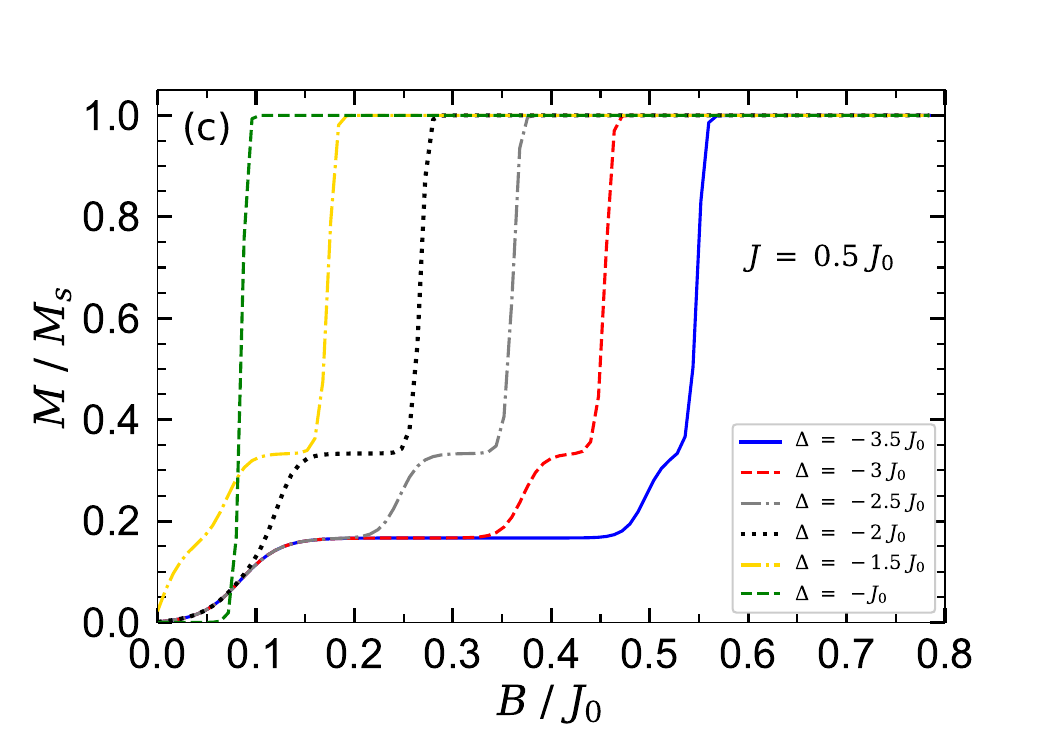}
}
\resizebox{0.45\textwidth}{!}{%
\includegraphics{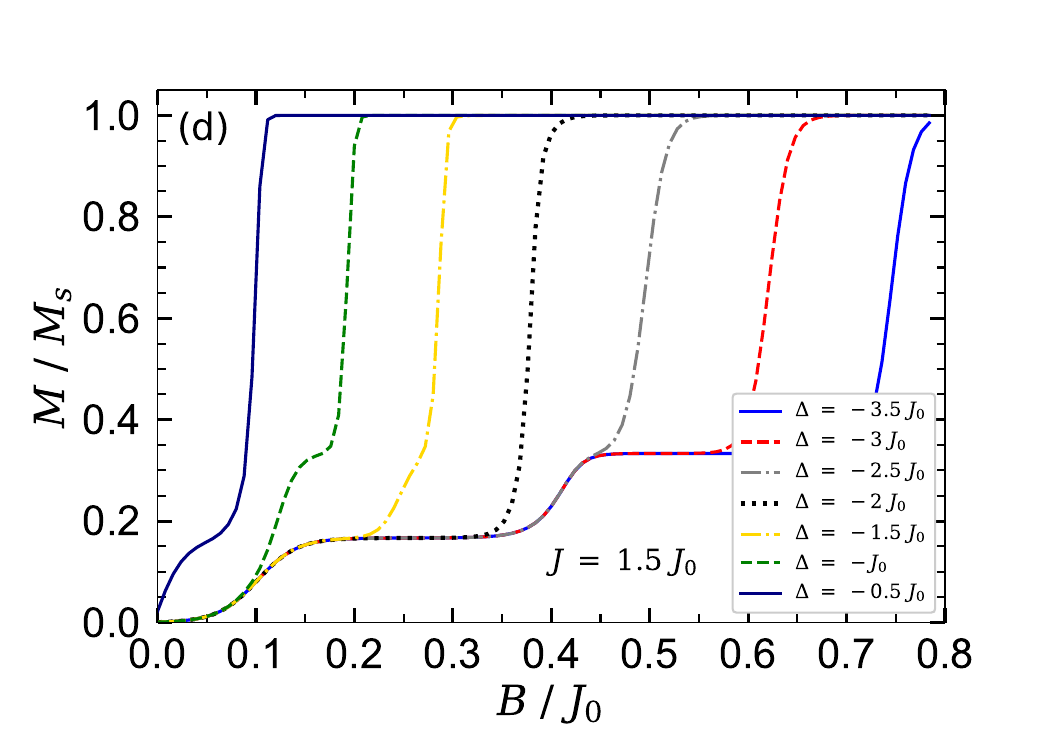}
}
\resizebox{0.45\textwidth}{!}{%
\includegraphics{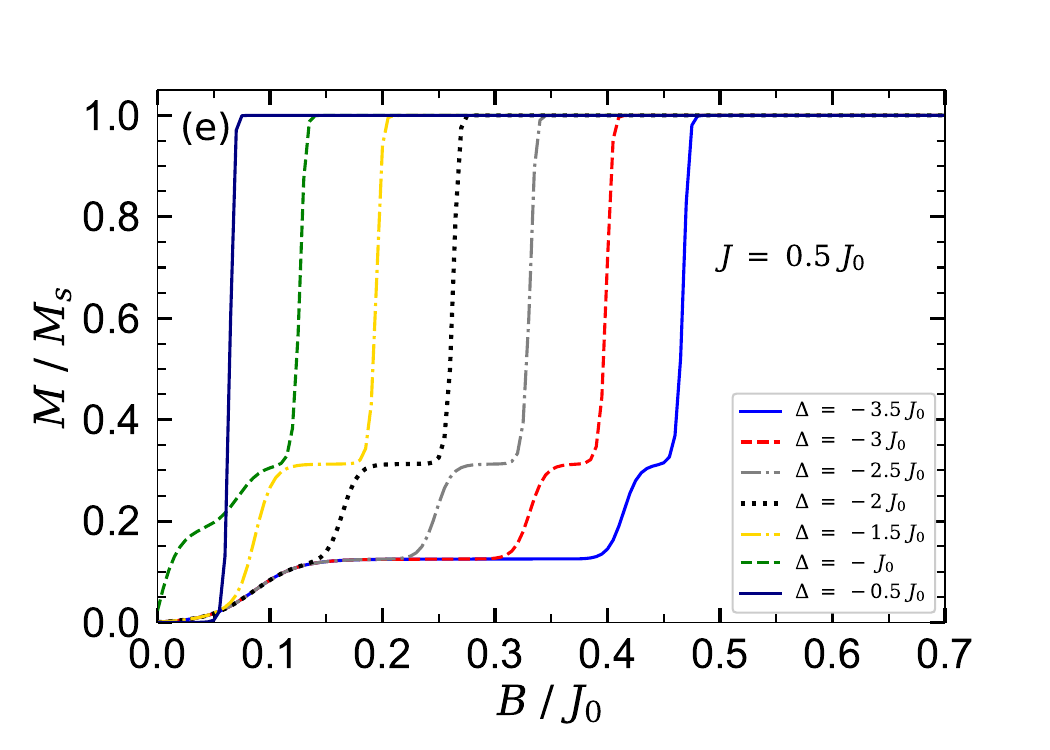}
}
\resizebox{0.45\textwidth}{!}{%
\includegraphics{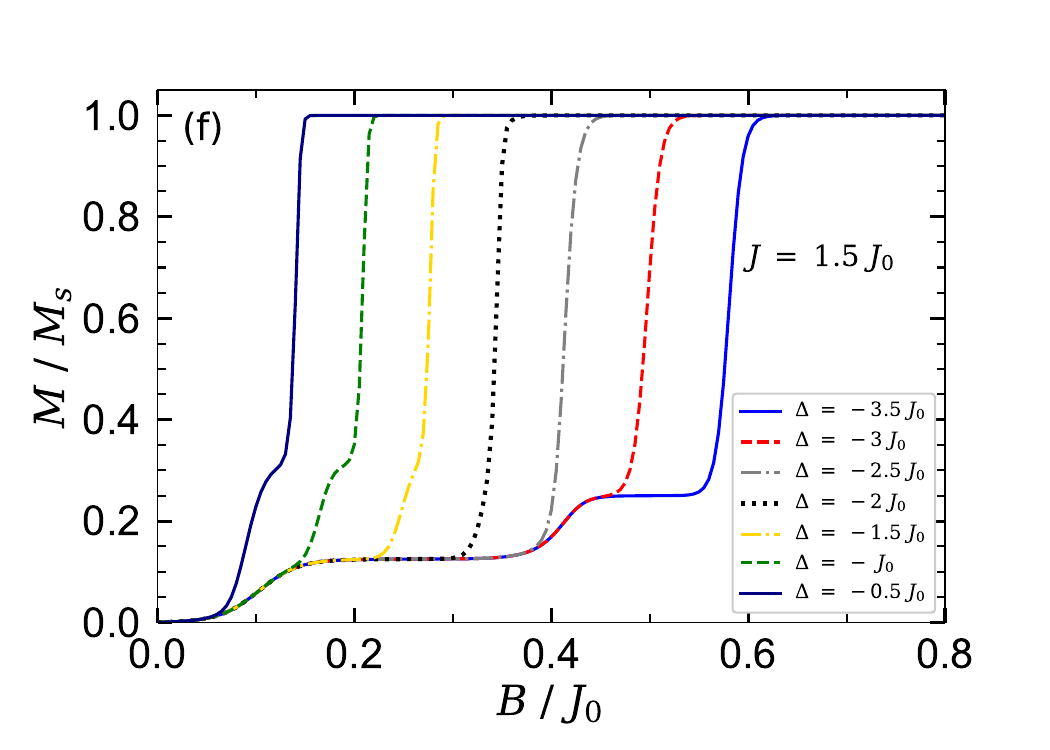}
}
\caption{(a) Magnetization in units of the saturation value $M/M_s$
  %of the spin-1/2 Ising-Heisenberg Cairo pentagonal model
  as a function of external magnetic field $B/J_0$ at temperature $T=0.02J_0$
  for $J=0.5J_0$, $g_1=g_2=1$ and several exchange anisotropy ratios 
  $\Delta/J_0$. (b) $J=1.5J_0$, where other parameters are taken as in panel (a).
  Magnetization for the cases (c) $J=0.5J_0$, $g_1=1$, $g_2=2$; 
  (d) $J=1.5J_0$, $g_1=1$, $g_2=2$; 
  (e) $J=0.5J_0$, $g_1=1$, $g_2=3$; and (f) $J=1.5J_0$, $g_1=1$, $g_2=3$.}
\label{fig:Mat}
\end{center}
\end{figure*}

Fig. \ref{fig:Mat} illustrates the magnetic field dependences of the magnetization per block
in the unit of its saturation value for various fixed values of the exchange anisotropy
parameter $\Delta/J_{0}$. Panel \ref{fig:Mat}(a) demonstrates the magnetization per block
against the magnetic field at low temperature ($T=0.02J_0$) and
fixed value of the isotropic coupling constant $J=0.5J_0$ and $g_1=g_2=1$,
where several values of the anisotropy $\Delta/J_{0}$ have been considered.
On the other hand, panel \ref{fig:Mat}(b) displays this quantity versus
the magnetic field at a larger coupling constant, $J=1.5J_0$,
for the same set of other parameters of the panel \ref{fig:Mat} (a). 
As mentioned before, the magnetization curve shows intermediate plateaux at zero,
one-eighth, one-fourth, three-eighth and one-half of saturation magnetization
for anisotropy range $\Delta/J_{0}<0$. It is shown in these figures
the magnetization jumps accompanied with the first-order ground-state
phase transition between different magnetization plateaux.
% from $0-$plateau to one-fourth plateau; from one-fourth plateau to one-half plateau and in turn from one-half plateau to the FP.
Critical magnetic fields at which magnetization jumps occur are observable as well. These
points strongly depend on both parameters $J/J_{0}$ and $\Delta/J_{0}$.

It is evident from Fig. \ref{fig:Mat}(b) that by tuning $J/J_0$ the magnetization jump occurs
at different critical magnetic fields. From the discontinuous ground-state phase transition
perspective, when the coupling constant $J/J_0$ increases, the transition from
the state with magnetization $M/M_s=1/4$ to that of with $M/M_s=1/2$
occurs at lower magnetic field, while the transition from ground-state
with the related magnetization $M/M_s=1/2$ to the FPS occurs at stronger magnetic fields
compared to the case $J=0.5J_0$. 

\begin{figure*}
\begin{center}
\resizebox{0.45\textwidth}{!}{%
\includegraphics[trim=30 5 30 5, clip]{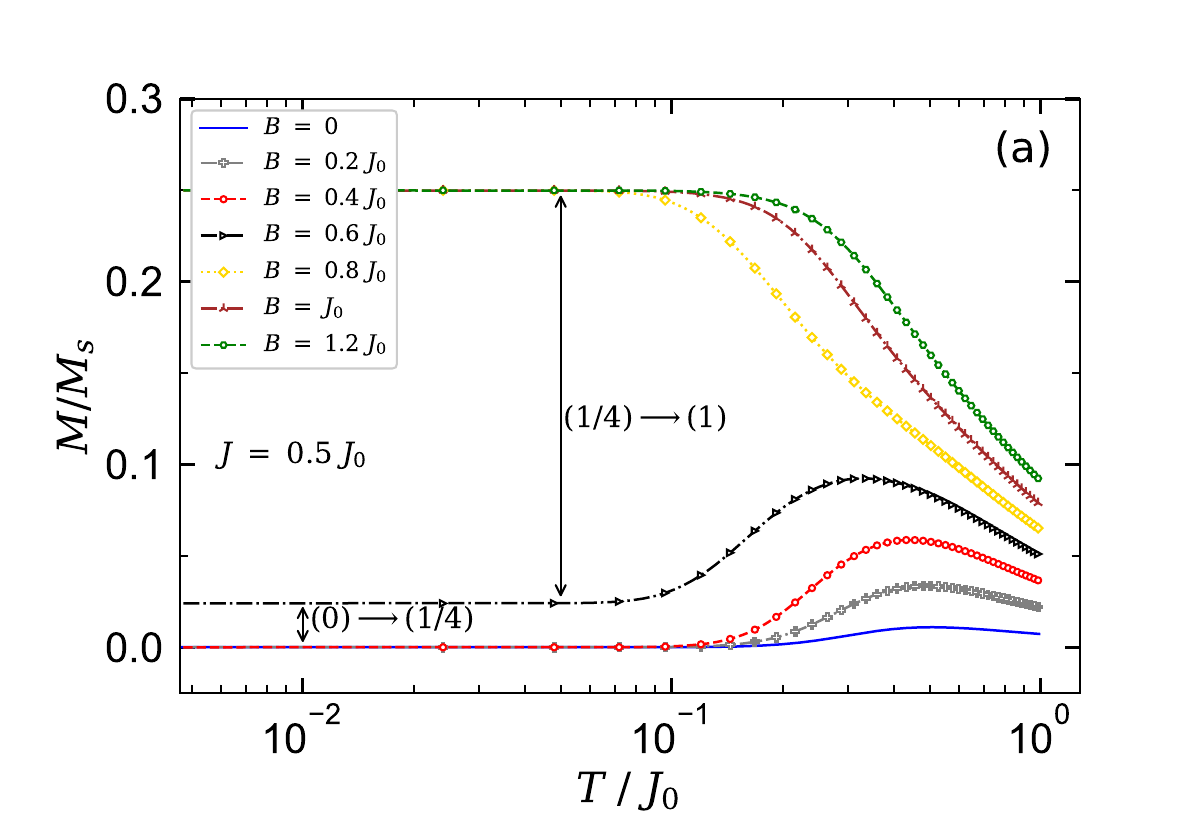}
}
\resizebox{0.45\textwidth}{!}{%
\includegraphics[trim=30 5 30 5, clip]{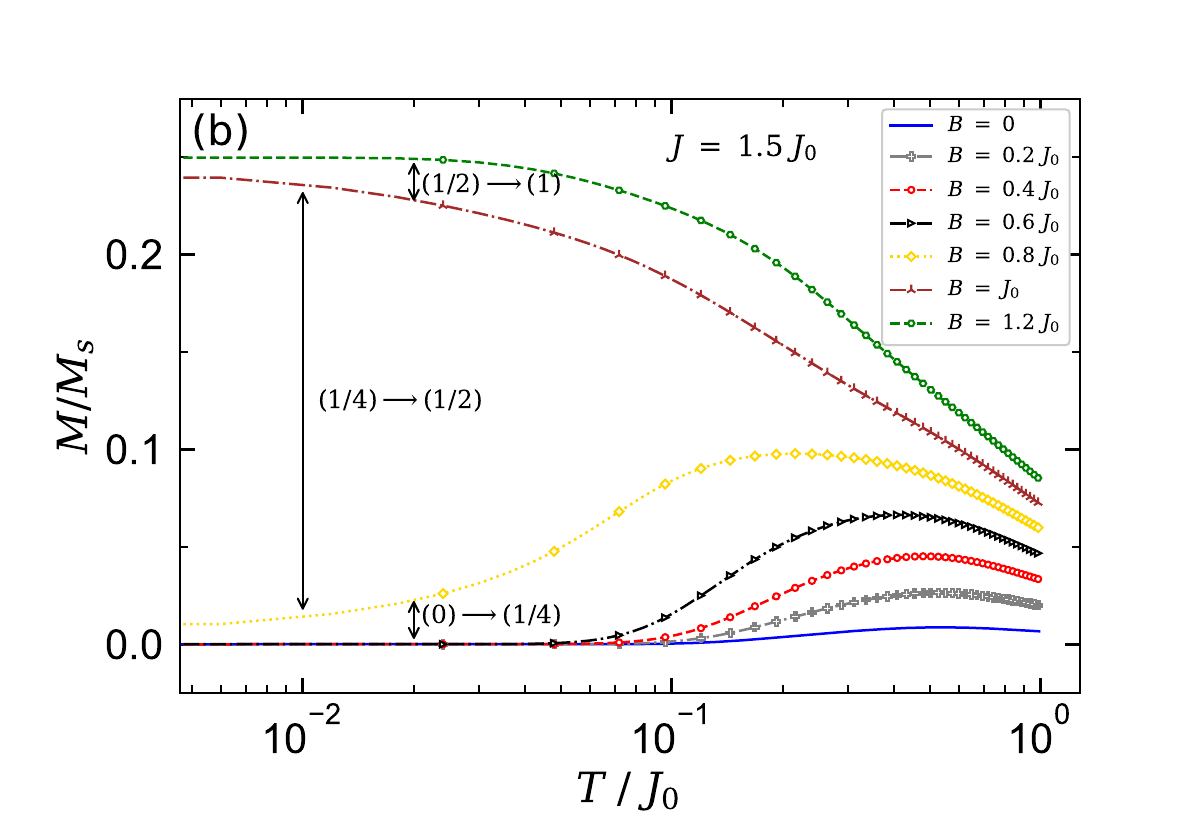}
}
\resizebox{0.45\textwidth}{!}{%
\includegraphics[trim=30 5 30 5, clip]{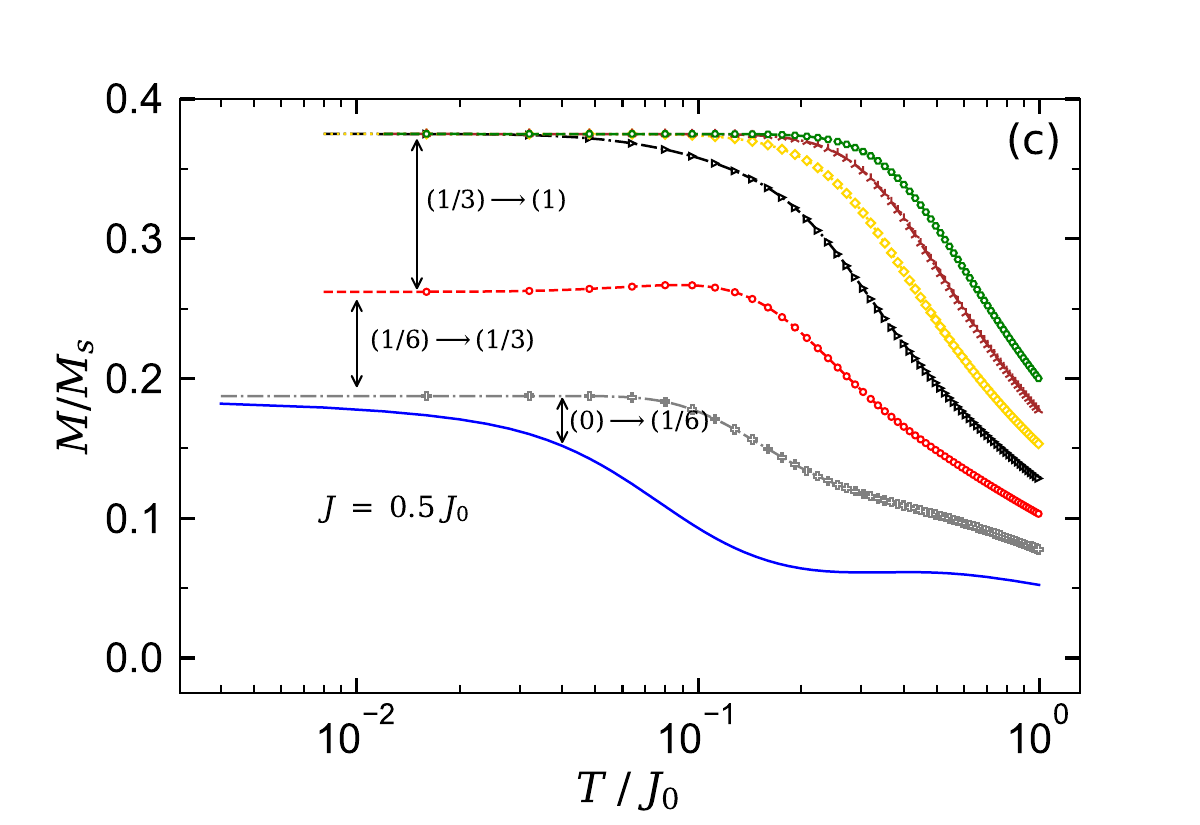}
}
\resizebox{0.45\textwidth}{!}{%
\includegraphics[trim=30 5 30 5, clip]{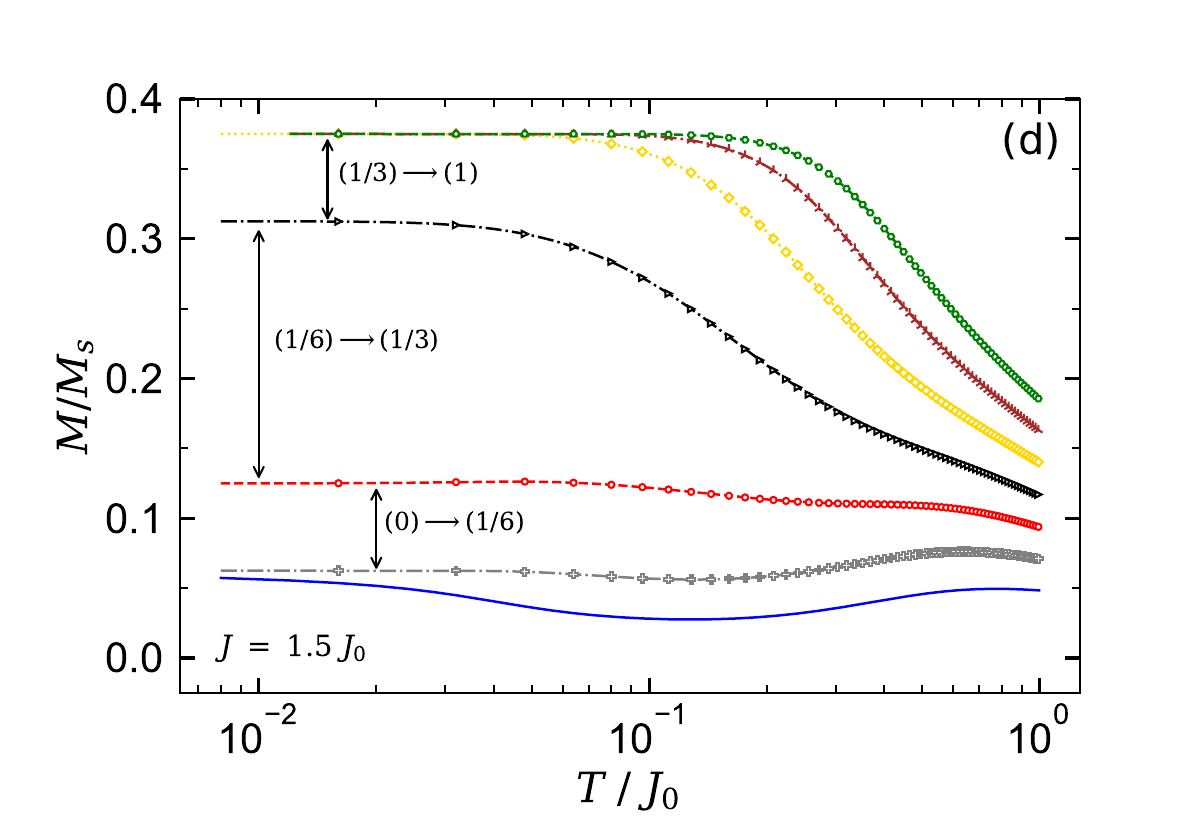}
}
\caption{Temperature dependence of the magnetization $M/M_s$
  %of the spin-1/2 Ising-Heisenberg Cairo pentagonal model
  for several selected magnetic fields and fixed $\Delta =-1.5J_0$,
  assuming (a) $J=0.5J_0$ and $g_1=g_2=1$; (b) $J=1.5J_0$ and $g_1=g_2=1$;
  (c) $J=0.5J_0$, $g_1=1$ and $g_2=2$; and (d) $J=1.5J_0$, $g_1=1$ and $g_2=2$.}
\label{fig:Mat_T}
\end{center}
\end{figure*}

Again assuming different g-factors $g_1=1$ and $g_2=2$,
we observe significant changes in the magnetization curve of the model. For instance, 
as shown in Fig. \ref{fig:Mat}(c), when fixed values $J=0.5J_0$, $g_1=1$ and $g_2=2$ are
considered, intermediate plateaux at zero, one-sixth and one-third of saturation magnetization
appear in the magnetization curve. By increasing the coupling constant $J/J_0$,
see Fig. \ref{fig:Mat}(d), the width of one-sixth plateau decreases,
while the one-third plateau gets wider so that the magnetization jump between
these plateaux occurs at a lower magnetic field.
% We now verify the effects of the temperature on the magnetization process of the model.

Fig. \ref{fig:Mat_T}(a) displays the temperature dependence of the magnetization
for several fixed values of the magnetic field for $J=0.5J_0$, $g_1=g_2=1$ and
at a fixed $\Delta =-1.5J_0$. The main effect of the temperature on the magnetization
is to aggregate the magnetization curves which are strictly discontinuous
at finite low temperatures. The magnetization gaps in this figure denote
the magnetization jump from one plateau to another.
Numbers in parentheses indicate the magnetization plateaux normalized
with respect to its saturation value. For example, $(0)\longrightarrow (1/4)$
means jumping from the zero plateau to the one-fourth plateau in the magnetization curve.
The magnetization gaps corresponding to the jumps from one-fourth to one-half and from
one-half to the saturation magnetization are quite evident in Fig. \ref{fig:Mat_T}(b).

The results illustrated in Figs. \ref{fig:Mat_T}(c) and \ref{fig:Mat_T}(d)
highlight %serve in evidence
that the magnetization behavior against temperature for the case $g_1=1$ and $g_2=2$
is substantially different from the case $g_1=g_2=1$.

{For the more general case $g_1=1$ and $g_2=ng_1$} with $n=1,\; 2,\; 3,\; \cdots$,
each magnetization plateau can be %pleasantly
successfully identified in terms of the considered 
Land{\' e} g-factors. %We derived the following relations:
By inspection we find
\begin{equation}
\label{MP}
\begin{array}{lcl}
\begin{cases}
\text{$g_1=1$}\\
 \text{$g_2=1$}\\
\end{cases}
\Rightarrow
M/M_s=\begin{cases}
 \text{$\frac{2}{8}$}\\
 \text{$\frac{3}{8}$}\\
  \text{$\frac{4}{8}$}\\
   \text{$\frac{1}{8}$}
\end{cases}
\equiv
\begin{cases}
 \text{$\dfrac{\frac{\alpha}{2}g_1}{\alpha(g_1+g_2)}$}\\
 \text{$\dfrac{\frac{\alpha}{2}g_1+g_2}{\alpha(g_1+g_2)}$}\\
  \text{$\dfrac{2(g_1+g_2)}{\alpha(g_1+g_2)}$}\\
   \text{$\dfrac{g_2}{\alpha(g_1+g_2)}$}
\end{cases},\\
\\
\begin{cases}
\text{$g_1=1$}\\
 \text{$g_2=2$}\\
\end{cases}
\Rightarrow
M/M_s=\begin{cases}
 \text{$\frac{1}{6}$}\\
 \text{$\frac{2}{6}$}\\
\end{cases}
\equiv
\begin{cases}
 \text{$\dfrac{\frac{\alpha}{2}g_1}{\alpha(g_1+g_2)}$}\\
 \text{$\dfrac{g_2}{\alpha(g_1+g_2)}$}\\
\end{cases}, \\
\\
\begin{cases}
\text{$g_1=1$}\\
 \text{$g_2=3$}\\
\end{cases}
\Rightarrow
M/M_s=\begin{cases}
 \text{$\frac{1}{8}$}\\
 \text{$\frac{5}{16}$}\\
 \text{$\frac{3}{16}$}
\end{cases}
\equiv
\begin{cases}
 \text{$\dfrac{\frac{\alpha}{2}g_1}{\alpha(g_1+g_2)}$}\\
 \text{$\dfrac{\frac{\alpha}{2}g_1+g_2}{\alpha(g_1+g_2)}$}\\
 \text{$\dfrac{g_2}{\alpha(g_1+g_2)}$}
\end{cases};
\end{array}
\end{equation}
{similar expressions could be carried out by increasing $n$.
  In Eqs. (\ref{MP}) it is}
{$$\alpha=2+[1+(-1)^{g_1+g_2}].$$}
%In above, we see general relations $M/M_s=\dfrac{\frac{\alpha}{2}g_1}{\alpha(g_1+g_2)}$,  $M/M_s=\dfrac{g_2}{\alpha(g_1+g_2)}$ and $M/M_s=\dfrac{\frac{\alpha}{2}g_1+g_2}{\alpha(g_1+g_2)}$ that repeat for each set of g-factors.
Figs. \ref{fig:Mat}(e) and \ref{fig:Mat}(f) %prove our claim
illustrate the validity of the results (\ref{MP}). E.g., 
in Fig. \ref{fig:Mat}(e) it is shown the magnetization per saturation
against the magnetic field for the situation $g_1=1$ and $g_2=3$,
where $J=0.5J_0$. It is
clear that there are magnetization intermediate plateuax at $M/M_s=1/8$, $M/M_s=5/16$ and
$M/M_s=3/16$, in agreement with the corresponding relationships noted in Eq. (\ref{MP}).

\subsection{Entropy and internal energy}

The magnetic field variations of the magnetization near the ground-state phase boundaries
may manifest themselves also in unusual behaviors of basic thermodynamic quantities,
%with respect to the temperature,
so we hereby explore in the following the temperature dependence %thermal variations
of the entropy and internal energy for different magnetic fields.

\begin{figure}
\begin{center}
\resizebox{0.7\textwidth}{!}{%
\includegraphics{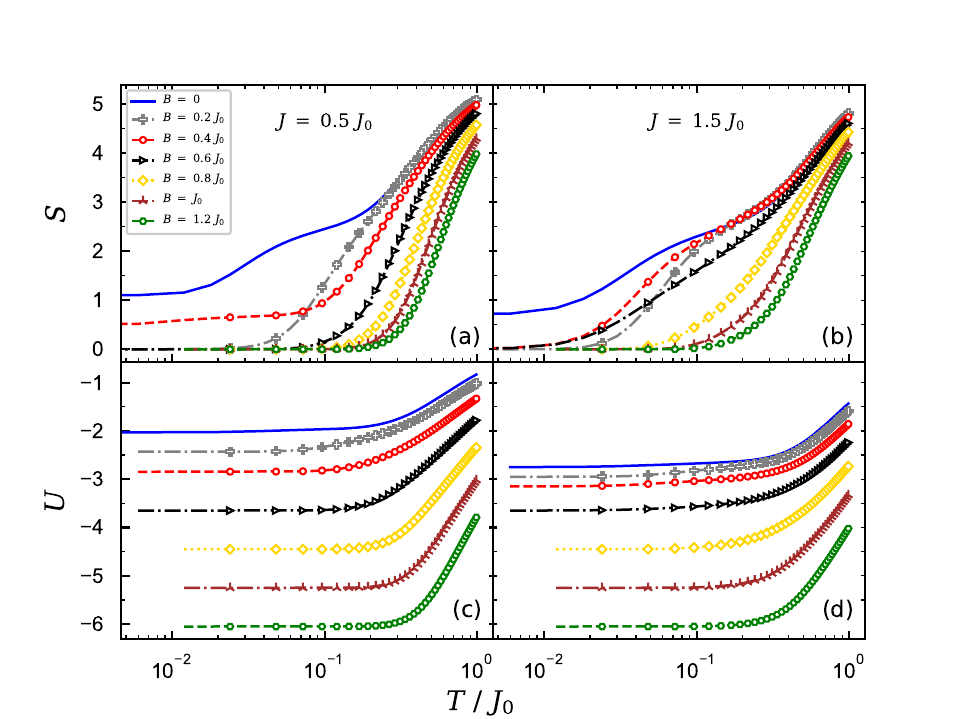}
}
\caption{(a) Entropy of the spin-1/2 Ising-Heisenberg Cairo pentagonal model
  as a function of the temperature for several fixed values of the magnetic field $B/J_{0}$ at
  fixed coupling constant $J=0.5J_{0}$. (b) The same as in panel (a), but with
  $J=1.5J_{0}$. Panels (c) and (d) illustrate the temperature dependence of the
  internal energy %of the model
  for the same selected magnetic fields and fixed values of other parameters as in
  panels (a) and (b), respectively. In all panels it is assumed
  fixed anisotropy $\Delta=-1.5J_{0}$ and %the case
  $g_1=g_2=1$. %has been taken
  %as a basic condition for the Land{\' e} g-factors.
}
\label{fig:EntropyInternalE}
\end{center}
\end{figure}

Figs. \ref{fig:EntropyInternalE}(a) and \ref{fig:EntropyInternalE}(b)
%demonstrate
show thermal variations of the entropy at different magnetic fields on a logarithmic scale
for $\Delta=-1.5J_{0}$, assuming $J=0.5J_{0}$ and $J=1.5J_{0}$, respectively.
As argued in Ref. \cite{Rodrigues} in the absence of the magnetic field, the
residual entropy at  $J=0.5J_{0}$ is given by $S \longrightarrow\mathrm{ln}(3)$
when $T\longrightarrow 0$. By applying a weak magnetic field
%($0<B< 0.2J_{0}$)
{($0<B \lesssim 0.2J_{0}$)}
the residual entropy at $J=0.5J_{0}$ decreases such that
$S< \mathrm{ln}(3)$ when $T\longrightarrow 0$. When the magnetic field
becomes {larger than $\approx 0.2J_{0}$},
the system is highly influenced by magnetization
one-fourth and one-half plateaux with residual entropy $S \longrightarrow 0$
when $T\longrightarrow 0$. For the case $J=1.5J_{0}$, even if a weak magnetic field is applied,
the system is dominated by ground-state phases associated to the one-fourth and one-half plateaux
with residual entropy $S \longrightarrow 0$ when $T\longrightarrow 0$.
Figs. \ref{fig:EntropyInternalE} (c) and \ref{fig:EntropyInternalE} (d)
display the internal energy $U = f + TS$, for the same set of parameters considered in
Figs. \ref{fig:EntropyInternalE} (a) and  \ref{fig:EntropyInternalE} (b), respectively.
Obviously, by applying an external magnetic field the internal energy
remarkably decreases at low temperatures. 

\begin{figure}
\begin{center}
\resizebox{0.7\textwidth}{!}{%
\includegraphics{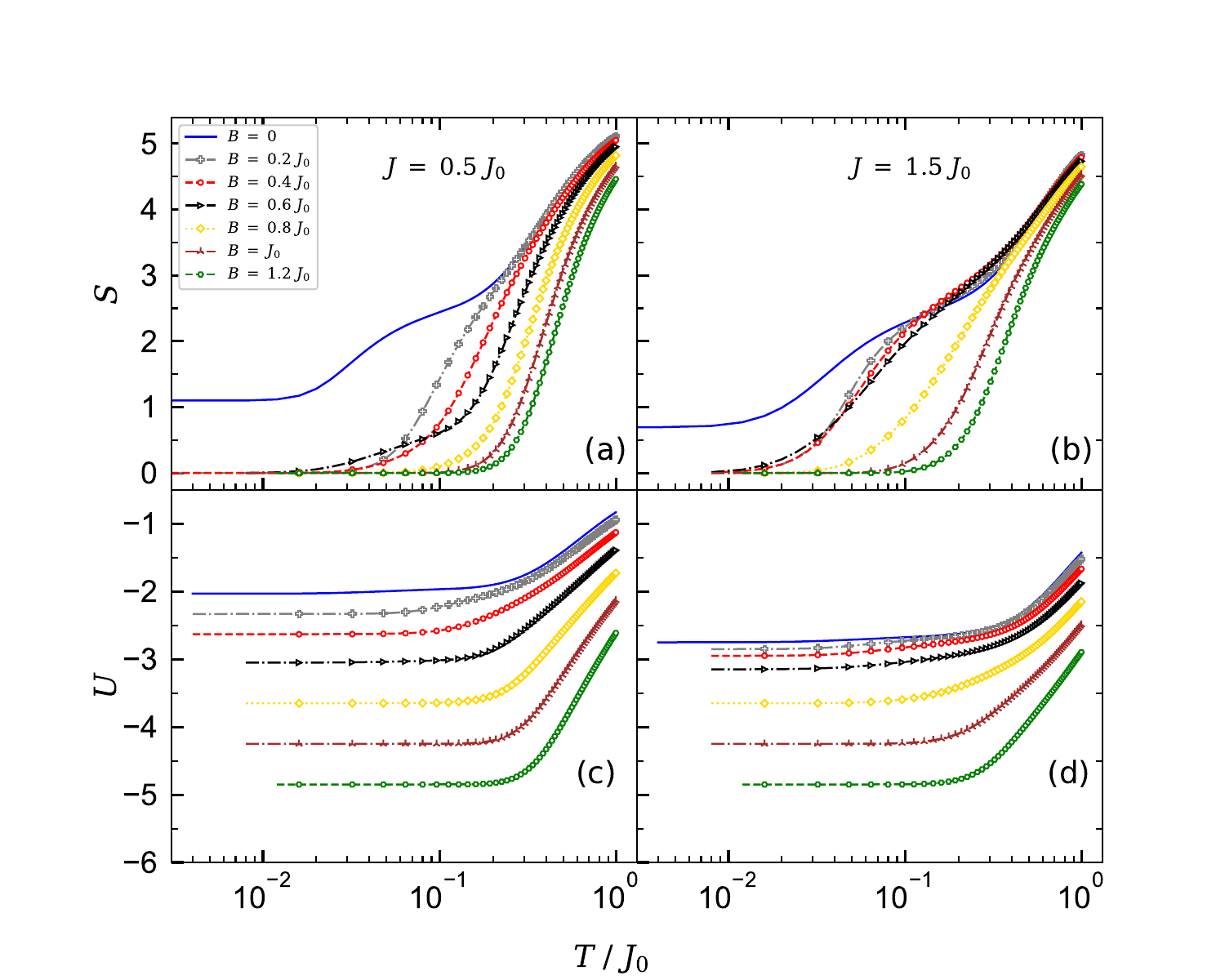}
}
\caption{(a) Entropy of the spin-1/2 Ising-Heisenberg Cairo pentagonal model
  as a function of the the ratio  $T/J_{0}$ for several
  fixed magnetic fields and $J=0.5J_{0}$. (b) The same as in panel (a),
  but with $J=1.5J_{0}$. In (c) and (d) we plot the temperature dependence
  of the internal energy of the model for the same selected magnetic fields and
  the same fixed values of other parameters as in panels (a) and (b), respectively. 
In all panels ii is $\Delta=-1.5J_{0}$, and $g_1=1$ and $g_2=2$.}
\label{fig:EntropyInternalE2}
\end{center}
\end{figure}

Let consider now $g_1=1$ and $g_2=2$.
Figs. \ref{fig:EntropyInternalE2}(a) and \ref{fig:EntropyInternalE2}(b)
show the entropy as a function of the temperature on the logarithmic scale for
the case when these different g-factors are considered.
By inspecting Fig. \ref{fig:EntropyInternalE2}(a) and comparing it
with Fig. \ref{fig:EntropyInternalE}(a), one can observe a non-trivial difference
in the entropy behavior versus temperature when an external magnetic field is applied.
For more clarity, if we focus on the three selected magnetic fields
$B=\{0.2J_0,\;0.4J_0,\;0.6J_0\}$ we realize that, independent of the ratio $J/J_{0}$,
the entropy behaves in a remarkably different way from the case $g_1=g_2=1$.
%Besides, change in the internal energy is sensible when we consider the situation  $g_1=1$ and $g_2=2$.
As depicted in Figs. \ref{fig:EntropyInternalE2}(c) and \ref{fig:EntropyInternalE2}(d),
the internal energy has been increased at low temperatures.

\subsection{Specific heat}

\begin{figure*}
\begin{center}
\resizebox{0.4\textwidth}{!}{%
\includegraphics{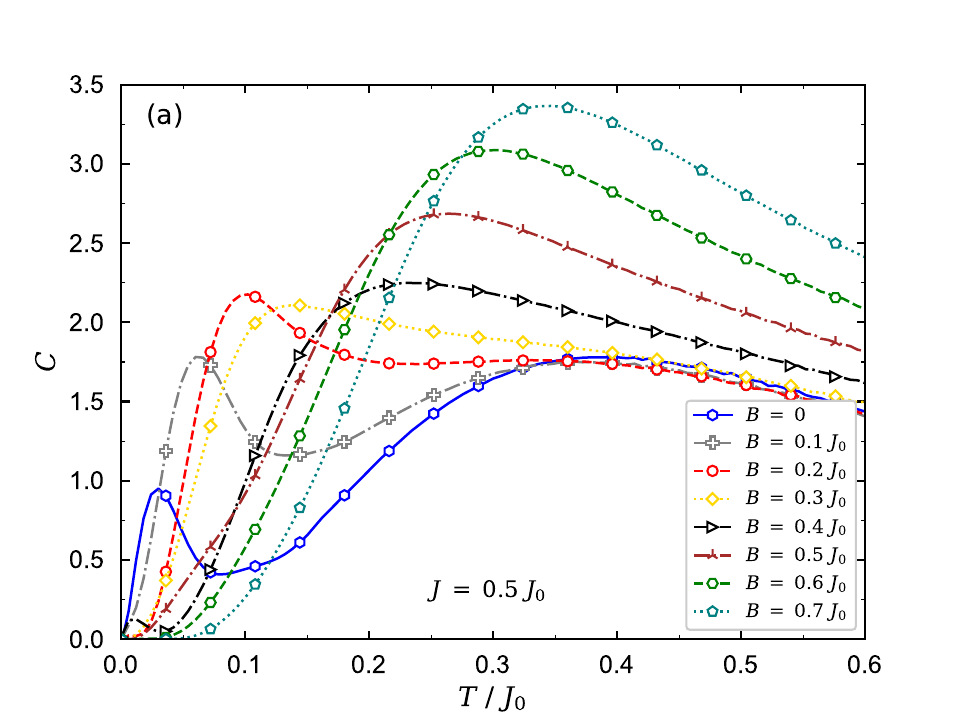}
}
\resizebox{0.4\textwidth}{!}{%
\includegraphics{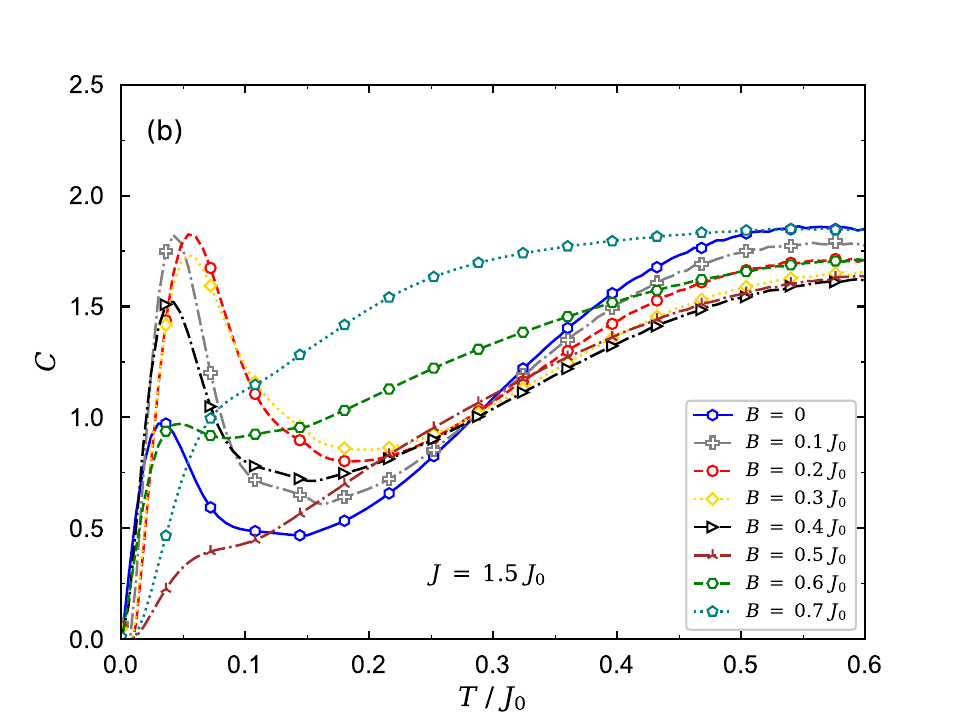}  
}
\resizebox{0.4\textwidth}{!}{%
\includegraphics{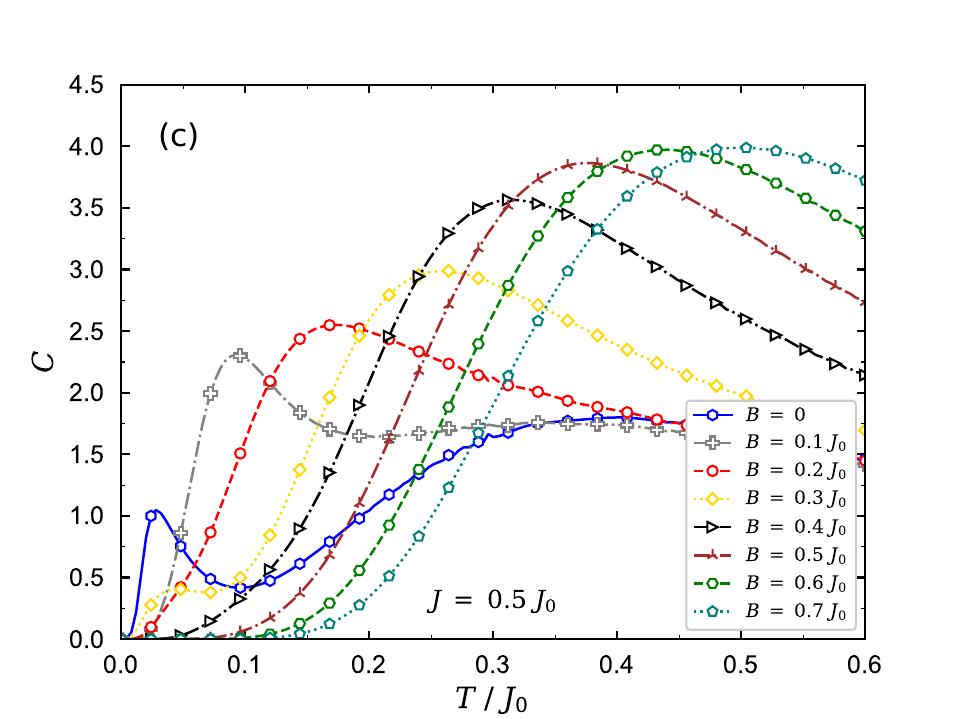}
}
\resizebox{0.4\textwidth}{!}{%
\includegraphics{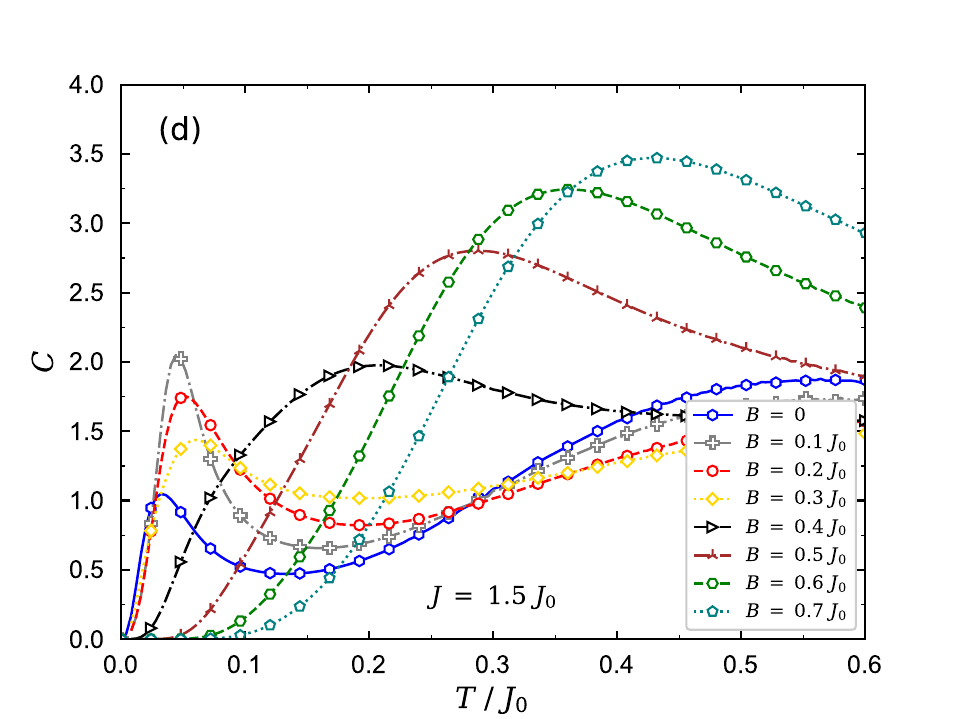} 
}
\caption{The specific heat of the spin-1/2 Ising-Heisenberg Cairo pentagonal model as
  a function of the temperature for several values of the magnetic field $B/J_{0}$:
  (a)$J=0.5J_{0}$, $\Delta=-1.5J_{0}$ and $g_1=g_2=1$; (b) $J=1.5J_{0}$, $\Delta=-1.5J_{0}$ and
  $g_1=g_2=1$; (c) $J=0.5J_{0}$, $\Delta=-1.5J_{0}$, $g_1=1$ and $g_2=2$; and (d)
  $J=1.5J_{0}$, $\Delta=-1.5J_{0}$, $g_1=1$ and $g_2=2$.}
\label{fig:SHeat}
\end{center}
\end{figure*}

Let us now examine the effects of exchange coupling $J/J_{0}$ and the magnetic field $B/J_{0}$
on the temperature dependence of the specific heat for anisotropy $\Delta=-1.5J_{0}$.
We display in Fig. \ref{fig:SHeat}(a) the specific heat of the model under consideration
as a function of the temperature for several fixed values of the magnetic field by supposing
$J=0.5J_{0}$ and  $g_1=g_2=1$. 
%This curve has similar behavior to the specific heat curve plotted in Ref. \cite{Rodrigues}. 
The blue solid line marked with hexagons represents the specific heat curve
for $J=0.5J_{0}$ and $\Delta=-1.5J_{0}$ when $B=0$. 
In Fig. \ref{fig:SHeat} (a) one can see that the specific heat
exhibits a double-peak when $B=0$. When the magnetic field increases,
two maxima of the double-peak gradually merge together and
make a single peak at higher temperature,
see for instance the red dashed line marked with cycles corresponding to $B=0.2J_0$.
With a further increase of the magnetic field %($B>0.2J_0$)
{($B \gtrsim 0.2J_0$)},
an anomalous Schottky type maximum arises at higher temperatures
(green dashed line marked with hexagons). 
%Interestingly, in the magnetic field interval $0.6J_0\lesssim B\lesssim 0.8J_0$, the smallest peak arises again at finite low temperatures. So, in this condition the specific heat again has a triple-peak temperature dependence. 
Comparison between Figs. \ref{fig:SHeat} (a) and \ref{fig:Mat} (a) %manifest
shows that the existence of the double-peak in the specific heat curve
denotes the magnetization zero plateau in magnetization curve. Therefore,
one can conclude that the spin-1/2 Ising-Heisenberg Cairo pentagonal model
is in the AFM state when a double-peak is observed in the specific heat curve.
The appearence of a single Schottky-peak nearby the critical magnetic field
$B\approx 0.3J_0$ (when $M/M_s=1$) indicates that the system is in the FPS.
Remarkably, a small change in the height of the first peak at low temperature, $T\approx 0.1J_0$,
is accompanied with the magnetization jump from the zero plateau to the one-fourth plateau,
see the evolution of specific heat from the red dashed line marked with cycles
to the gold dotted line marked with diamonds illustrated in Fig. \ref{fig:SHeat}(a).
In fact, the height of the first peak gradually increases as the magnetic field increases,
then slightly decreases for the magnetic field interval %$0.2J_0<B<0.3J_0$.
{$0.2J_0 \lesssim B \lesssim 0.3J_0$}.

 % With further increase of the magnetic field, the specific heat has a temporary double-peak. Surprisingly, the intermediate magnetization $1/2-$plateau coincides the appearance of this double-peak.
% In the presence of strong magnetic fields, the double-peak merge together and make a permanent single Schottky peak. This phenomenon testifies the model is in the CFM phase (trace the evolution of specific heat curve from marked green dotted dash line to the blue dashed line marked with cycles illustrated in Fig. \ref{fig:SHeat}(a)).
%Ultimately, the magnetization jump from $1/2-$plateau to the saturation magnetization is in coincidence with the arising of a small peak when the magnetic field increases further than $B=J_0$ .

To gain  further insight into the effects of the coupling constant $J/J_{0}$
on the specific heat, let us also examine the specific heat behavior of
the spin-1/2 Ising-Heisenberg Cairo pentagonal model for different values of $J/J_{0}$,
which can be of particularly interest especially
when the model is in the presence of a tunable magnetic field.
We have depicted in Fig. \ref{fig:SHeat}(b) the typical dependences of the specific
heat on the temperature  or several magnetic fields, assuming
$J=1.5J_{0}$, $\Delta=-1.5J_{0}$ and $g_1=g_2=1$. The interesting point
from this figure is that the specific heat does not show the single Schottky-peak
for range
%$B<0.7J_0$
{$B \lesssim 0.7J_0$}.
The particular double-peak temperature dependence
is observed whose peaks discontinuously change in height upon increasing the magnetic field.
The rise and fall of the height of first peak appeared at lower temperatures is accompanied
with the presence of magnetization plateaux. 

Finally, the specific heat has a Schottky-type maximum at higher temperature and
for large magnetic fields %($B>0.7J_0$)
{$B \gtrsim 0.7J_0$}, revealing the model is in a FPS.
Another interesting thing is that the Schottky-type maximum appears for 
magnetic fields quite larger than the case $J=0.5J_{0}$. This phenomenon
indicates the existence of a magnetization intermediate plateau at one-half
of saturation magnetization for the case  $J=1.5J_{0}$. Consequently, when the coupling
constant $J/J_{0}$ take larger values, the single Schottky peak occurs for
magnetic fields larger than the ones for $J=0.5J_{0}$. This scenario is in
agreement with the width alterations of intermediate $1/2-$plateau, as
shown in Fig. \ref{fig:QPT}(b). 

For completeness, Fig. \ref{fig:SHeat}(c) shows the specific heat versus temperature
for a number of selected magnetic fields,
assuming $\Delta=-1.5J_{0}$, $J=0.5J_{0}$, $g_1=1$ and $g_2=2$. It is clear that the
Schottky maximum appears at lower magnetic field and higher temperature intervals.
By increasing $J/J_{0}$, as reported in Fig. \ref{fig:SHeat}(d),
the distance between the peaks of the double peak structure
increases and the single maximum arises at considerably lower magnetic fields
($B\gtrsim 0.4J_{0}$) with respect to the case when $g_1=g_2=1$.

\section{Conclusions}\label{conclusions}

In summary, we have %theoretically
investigated the thermodynamic and magnetic properties
%=properties and magnetic behavior
of the spin-1/2 Ising-Heisenberg
Cairo pentagonal model in the presence of an external tunable magnetic field.
%As a matter of fact,
The magnetization process, entropy, internal energy and the specific heat of this model have
been %widely investigated
studied by means of the solution obtained within the transfer-matrix formalism.
The possibility of having different
%situations for the
Land{\'e} g-factors was considered.

% In terms of numerical investigations, we understood
We showed that the magnetization %has
exhbits intermediate plateaux
at zero, one-fourth and one-half of the saturation magnetization
when the same Land{\'e} g-factors are assumed for the nodal Ising spins and dimer spins.
The width of these plateaux %rigorously
depends on the both the isotropic and anisotropic Heisenberg exchange couplings
considered for the dimers. %In fact,
By decreasing the anisotropic exchange coupling, the magnetization intermediate
one-fourth and one-half plateaux gradually appear. Increasing
the isotropic coupling constant results in widening the width of one-half plateau and
decreasing %limiting
the width of one-fourth one. The magnetic field variations considerably affect
the entropy and the internal energy when the temperature goes to zero.
When different g-factors are considered, %for the nodal spins and dimer spins results in appearing
different intermediate plateaux are obtained. 
%at one-sixth and one-third of saturation value .
We also observed a substantial change in entropy behavior and internal energy of the model.
We focused on the case in which the ratio between the Land{\'e} g-factors
  is an integer, and
  it would be interesting to consider in the future both the more general
  case in which the ratio is a rational number and the situation
  in which the g-factors are incommensurable.
 
It has been already shown that in the absence of magnetic field 
the specific heat curve of the model manifests a visible double-peak structure
\cite{Rodrigues}. Here, we have shown that turning on and
varying the magnetic field and
the isotropic coupling constant can remarkably %alter
modify the shape and the temperature position of this double-peak. %Actually, under the magnetic field variations the height and temperature-position of the both peaks undergo significant changes.
One of he most %stimulating
interesting result we found is provided by the
%finding stemming from our study represents an outstanding diversity of
magnetic field dependence of the specific heat, for which
%in the specific heat curve all anomalous
the rises and falls of the first peak are %pleasantly
well in agreement %compatible
with the entity of magnetization plateaux.
%Hence, we claimed
We concluded that one can recognize the lowest-energy eigenstate of the
model by investigating the evolution of first peak of the double-peak.
With a further increase of the magnetic field, the first peak gradually disappears
and a single Schottky peak is created at higher temperatures.
This single Schottky peak %denotes
indicates that the model is in the fully polarized phase.  
As a result, varying the exchange interactions between dimer spins and
selecting different Land{\' e} g-factors have
notable effects on the specific heat behavior versus temperature. 
 %As a result, the specific heat variations with respect to the magnetic field are in a good accordance with the low-temperature magnetization process of the spin-1/2 Ising-Heisenberg Cairo pentagonal model.

Finally, we mention that it would be interesting to consider more general
  configurations in which spin-1/2 Ising-Heisenberg
  Cairo pentagonal chains are connected, starting from the case in which they are merged
  as a $Y$-junction wiht a variable leg, a geometry which has been considered
  for XX and Ising quantum spin models \cite{Crampe,Tsvelik,Giuliano}.

\section*{Acknowledgments}
The authors are grateful to O. Rojas and S. Ruffo for useful discussions.
H.A.Z. and N.A. acknowledge the receipt of the grant from the Abdus Salam International Centre for Theoretical Physics (ICTP), Trieste, Italy,
and the CS MES RA in the frame of the research project No. SCS 18T-1C155, and the CSMES RA in the frame of the research project No. SCS 19IT-008.
N.A. and A.T. acknowledge support from the CNR/MESRA project
 "Statistical Physics of Classical and Quantum NonLocal Hamiltonians: Phase Diagrams and Renormalization Group".

% \section*{References}

\bibliography{}

\end{document}